\documentclass[aps,pra,showpacs%
        ,amsmath%
        ,superscriptaddress%
        ,reprint%
        ,nofootinbib
        ,preprintnumbers
        ]{revtex4-1}
\usepackage[pdftex]{graphicx}

\usepackage{amssymb}
\usepackage{amsmath} 

\usepackage[mathscr]{eucal}

\newcommand{\pdfrac}[2]{\frac{\partial #1}{\partial #2}}

\newcommand{\bracket}[2]{\langle{}#1{}|{}#2{}\rangle}

\newcommand{\bvec}[1]{\boldsymbol{#1}} 

\begin{document}


\title{Generating nonequilibrium stationary state from ground state condensate
  \\
  through an almost-adiabatic cycle}


\author{Atushi Tanaka}
\homepage[]{\tt http://researchmap.jp/tanaka-atushi/}
\affiliation{Department of Physics, Tokyo Metropolitan University,
  Hachioji, Tokyo 192-0397, Japan}
\author{Takaaki Nakamura}
\affiliation{Laboratory of Physics, Kochi University of Technology,
  Tosa Yamada, Kochi 782-8502, Japan}
\author{Taksu Cheon}
\affiliation{Laboratory of Physics, Kochi University of Technology,
  Tosa Yamada, Kochi 782-8502, Japan}

\date{\today}

\begin{abstract}
It is shown that the ground state of weakly interacting Bose particles
in a quasi one-dimensional box trap can be converted into an excited
stationary state by an adiabatic cyclic operation that involves a
quench of interaction strength: A sharp impurity potential is applied
and its strength is varied during the cycle, which induces a
nonequilibrium stationary state exhibiting the inversion of
population. This process is robust in the sense that the resultant
stationary state is almost independent of the details of the cycle,
such as the position of the impurity as long as the cycle is far
enough from critical regions. The case of the failure of the
population inversion due to the strong interparticle interactions is
also examined.
\end{abstract}


\maketitle

\section{Introduction}
\label{sec:Introduction}

According to the principle of the equilibrium thermodynamics, a
quasi-static adiabatic cycle is trivial, in the sense that the initial
and final states are identical~\cite{Callen}.
Once we
slightly relieve the quasi-static adiabatic condition,
or
the
thermodynamic condition,
however,
a cyclic operation may transform a stationary
state into another one.
A promising system to realize such stationary state transformations in a
many-body, nearly thermodynamic setting is cold atoms. This is because
it is possible to manipulate the system with its
quantum coherence intact, as it can be well-isolated from
environmental degrees of freedom.

An example~\cite{Yonezawa-PRA-87-062113} can be found in the
Lieb-Liniger model~\cite{Lieb-PR-130-15,Guan-IJMPB-28-1430015}, which
describes Bose particles confined in a one-dimensional ring. A
stationary state of the Lieb-Liniger model is delivered to another
stationary state through a cyclic operation where the strength of
interparticle interaction is adiabatically increased except at a
point: Once the interparticle interaction becomes infinitely repulsive
to make the Tonks-Girardeau regime~\cite{Girardeau-JMP-1-516}, 
the interparticle interaction
strength
is
flipped
to infinitely attractive to
form the super-Tonks-Girardeau
regime~\cite{Astrakharchik-PRL-95-190407,Haller-Science-325-1224}. During
the cyclic operation, the stationary state is smoothly deformed even
at the
strength flipping
point, where normalizable stationary states are ensured
to be kept unchanged.

The state transformations in the Lieb-Liniger model is an example of exotic
quantum holonomy in adiabatic cycles of microscopic, non-thermal
systems: Although one may expect that
an adiabatic cycle
brings no change in
stationary states up to a phase factor,
it
may
transform a stationary state into
another,
for example,
in Floquet systems
through the winding of
quasienergy~\cite{Tanaka-PRL-98-160407,Kitagawa-PRB-82-235114,Zhou-PRB-94-075443}
and Hamiltonian systems with level
crossings~\cite{Cheon-PLA-374-144,Spurrier-PRA-97-043603}.
%
A similar, but distinct concept to the exotic quantum holonomy,
Wilczek-Zee's holonomy, where an adiabatic cycle offers a
transformation of degenerated stationary
states~\cite{Wilczek-PRL-52-2111,Bohm-GPQS-2003,Chruscinski-GPCQM-2004},
is
also
utilized to control quantum
states~\cite{Zanardi-PLA-264-94,Tanimura-PLA-325-199,Sjoeqvist-NJP-14-103035}.

We note that the adiabatic state transformation in the Lieb-Liniger
model~\cite{Yonezawa-PRA-87-062113} heavily depends on the
particularity
of
the model.
The number of
particles is required to be specified precisely. Also, the theoretical
argument in Ref.~\cite{Yonezawa-PRA-87-062113}
depends
essentially
on
the solvability of the
system.
Hence it seems
difficult to
extend this result to other quantum many-body systems.

In this paper, we examine
an
adiabatic state transformation in a
simple quantum many-body system, which might be extensible to various
ways.  We here examine cyclic operations on Bose particles confined in
a
quasi one-dimensional box
trap~\cite{Meyrath-PRA-71-041604,Batchelor-JPA-38-7787,%
  Syrwid-PRA-96-043602,Sciacca-PRA-95-013628}
to
generate a
nonequilibrium stationary state
from the ground state.
Applying an
almost%
-%
adiabatic cycle
that involves
a quench of the strength
of
a sharp impurity
potential,
we obtain a
population-inverted state, in which the Bosons occupy a single-particle
excited state.
The population inversion
has been
utilized to achieve negative
temperature~\cite{Purcell-PR-81-279}, for example, and is related to
dark solitons in the studies of Bose-Einstein
condensates~\cite{Dum-PRL-80-2972,Damski-PRA-65-013604}.

Our starting point is the analysis of cyclic operations for
noninteracting Bose
particles~\cite{Kasumie-PRA-93-042105,Tanaka-NJP-18-045023}.  We
define a cycle using a sharp impurity potential, which is often
utilized to manipulate condensates both
experimentally and
theoretically~\cite{Piazza-PRA-80-021601,Karkuszewski-PRA-63-061601,Frantzeskakis-PRA-66-053608,Hans-PRA-92-013627,Ramanathan-PRL-106-130401,Kunimi-PRA-100-063617}.
In order to incorporate the interparticle interaction, we suppose that
the system is described by the time-dependent Gross-Pitaevskii
equation. We
show that the interparticle interaction can
significantly modify the almost-adiabatic processes due to the
appearance of bifurcations in the solution of time-independent
Gross-Pitaevskii
equation
such as
the
swallowtail structure~\cite{Wu-PRA-61-023402,Damski-PRA-65-013604,%
Liu-PRA-66-023404,Mueller-PRA-66-063603,Kunimi-PRA-91-053608,%
PerezObiol-arXiv-1907.04574}.

The plan of this paper is the following. In Sec.~2, we introduce a
cyclic operation
that involves
a flip of the potential strength
which is analogous to
the cyclic operation
introduced in Ref.~\cite{Kasumie-PRA-93-042105} for non-interacting
systems.
In Sec.~3, we numerically examine the cyclic operation in the repulsively
interacting Bose
particles
using the Gross-Pitaevskii equation,
where it is shown that the
strong
nonlinearity
invalidates
the population
inversion.
In Sec.~4, a theoretical interpretation for the
numerical result
is shown.
Sec.~5 concludes this paper with
summary and outlook. Appendix~\ref{sec:Bogoliubov} offers
details of the linear stability analysis (the Bogoliubov analysis)
at the
quench
point.

\section{Cycle for non-interacting particles}
We look at a cyclic operation for particles confined in a
quasi one-dimensional
boxed trap
with
an impurity
potential. We
illustrate how this cycle works for non-interacting
particles~\cite{Kasumie-PRA-93-042105}, which offers a basis
for examining
the case of interacting Bosons.

We assume that a particle is described by the one-dimensional
time-dependent Schr\"odinger equation
\begin{equation}
  \label{eq:defTDS0}
  i\hbar\pdfrac{}{t}\Psi(x,t)
  = -\frac{\hbar^2}{2M}\pdfrac{^2}{x^2}\Psi(x,t)
\end{equation}
under the boundary condition $\Psi(0,t)=\Psi(L,t)=0$, where $M$ is the
mass of the particle, and $L$ is the size of the box trap.  In the
following, we assume $\hbar=1$, $M=1/2$ and $L=2\pi$.

After the system is prepared to be in a stationary state, a
sharp impurity potential
is placed at $x=X$ to realize cyclic operations, where the
strength $v$
is varied.
We assume that the
impurity
potential
$V_{\rm i}(x; v)$ is
described by the
Dirac's
delta function:
\begin{equation}
  \label{eq:defVi}
  V_{\rm i}(x; v) = v\delta\left(x - X\right)
  .
\end{equation}

There are two building blocks of the cyclic operation: One is the smooth
and monotonic variations of parameter: $C_{\rm s}(v', v'')$,
which denotes the variation of $v$ from $v'$ to $v''$,
while keeping the value of $X$.
The other is the discontinuous operation $C_{\rm d}(v',v'')$ in
which the value of $v$ is
changed
from $v'$ to $v''$, which resembles
the process that is utilized to create the super-Tonks-Girardeau gas
from the Tonks-Girardeau gas~\cite{Haller-Science-325-1224}.

\begin{figure}[hbt]
  \centering
  \includegraphics[%
  	height=3.5cm%
        ]{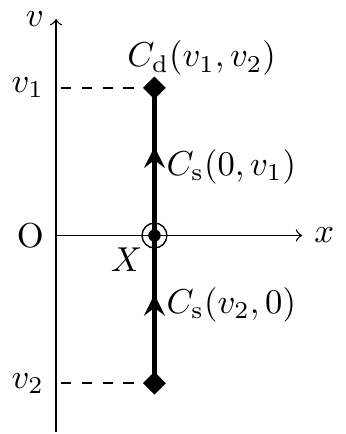}
  \caption{%
    The closed path $C(X)$ in $(x,v)$-plane. Throughout the cycle,
    the value of $X$ is fixed.
    At the initial point, we assume $v=0$.
    To close the cycle, $v_1=\infty$ and $v_2=-\infty$ are
    assumed. In numerical experiments, we assume that
    the values of $v_1$ and $-v_2$ are large, but finite.
  }
  \label{fig:paths_Xv}
\end{figure}

We define the almost-adiabatic cyclic operation $C(X)$
(see, Fig.~\ref{fig:paths_Xv}), which
involves a
quench
of the
impurity
potential placed
at $X$. This cycle is a succession of three operations
$C_{\rm s}(0, \infty)$, $C_{\rm d}(\infty, -\infty)$ and
$C_{\rm s}(-\infty, 0)$.

\begin{figure}[hbt]
  \centering
  \includegraphics[%
  	width=7.0cm%
        ]{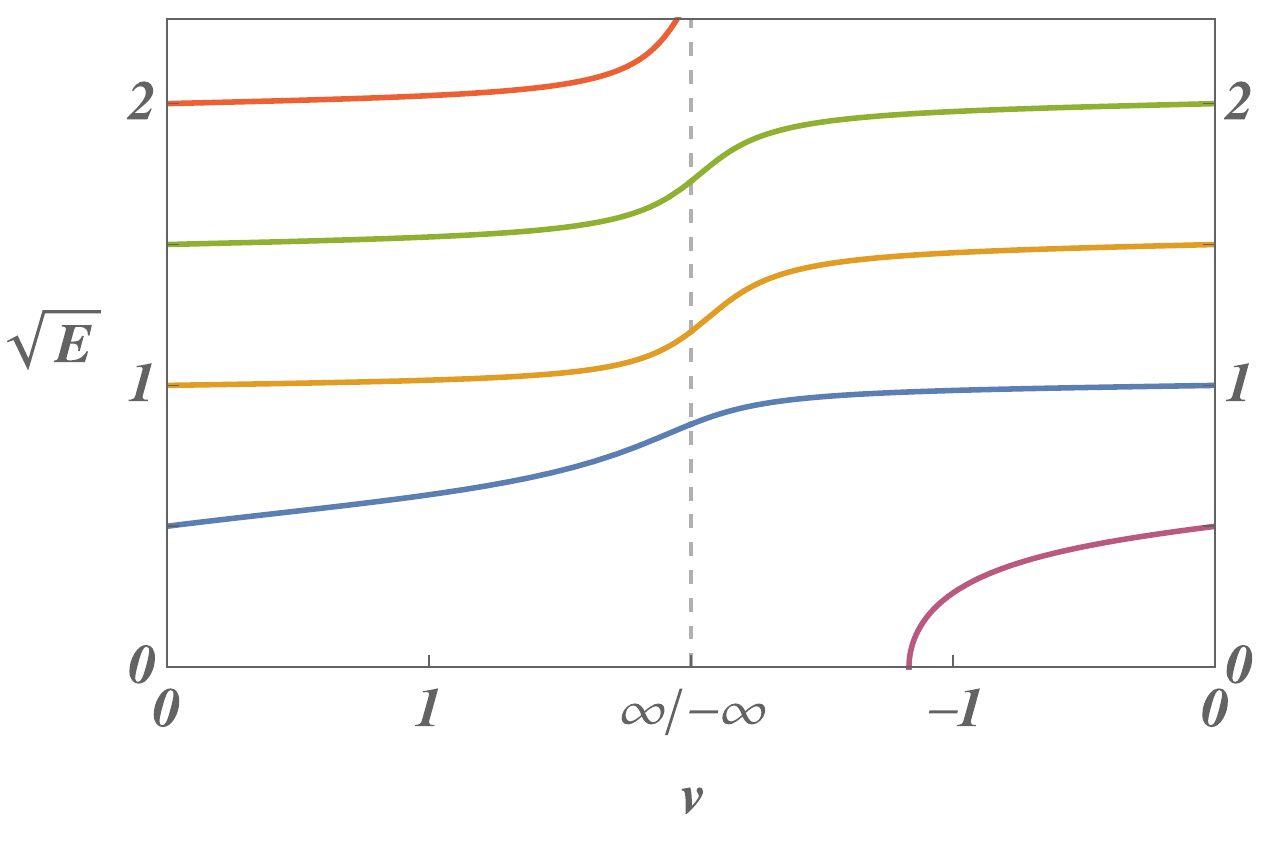}
  \caption{%
    (Color online) Parametric evolution of eigenenergies along
    $C(X=0.42L)$, which connects $n$-th and $(n+1)$-th eigenenergies. The
    horizontal axis is linear in $\tan^{-1} v$.
  }
  \label{fig:linearCq}
\end{figure}

We show that an initial stationary state $\Psi_n(x)$ can be
transformed
to another stationary state after the completion of a cycle.
Here we assume that the parameters are varied adiabatically during the smooth
operations.
Hence the system is governed by the adiabatic theorem~\cite{Kato-JPSJ-5-435}.
Since the relevant eigenenergy and eigenfunction are continuous during
the
quench,
the parametric dependence of eigenenergies tells us the final
stationary state for the almost-adiabatic cycle $C(X)$.
We depict an example of the parametric evolution of eigenenergies in
Fig.~\ref{fig:linearCq}.

After the completion of the almost-adiabatic cycle $C(X)$,
the final state is $\Psi_{n+1}(x)$, as long as
$X$ does not coincide with the node of the initial wavefunction
$\Psi_n(x)$. This is because the
eigenenergies are increased monotonically during
$C_{\rm s}(0, \infty)$ and $C_{\rm s}(-\infty, 0)$ as $v$ is
increased monotonically~\cite{Kasumie-PRA-93-042105,Simon-CPAM-39-75},
and are continuous at the quench process
$C_{\rm d}(\infty, -\infty)$~\cite{Kasumie-PRA-93-042105}.

Let us extend the above argument to the non-interacting Bose
particles. Suppose that all particles occupy the single-particle
ground state $\Psi_1(x)$. After the completion of the almost adiabatic
operation $C(X)$, all particles occupy the first excited state
$\Psi_2(x)$. Namely, a population
inverted
state can be created from the ground state through the almost
adiabatic operation $C(X)$.

Although the almost-adiabatic cycle $C(X)$ induce exotic changes
as shown above,
the cycle involves several idealizations.
In particular,
we need to take into account the effect of interparticle interaction.
In
the following sections, we scrutinize the almost-adiabatic cyclic operation $C(X)$ for interacting Bose particles.

\section{%
  Cycle for
  an interacting Bose system}
We here examine the almost-adiabatic cycle $C(X)$ in a many-body
setting.  We assume that Bose particles are confined in the
quasi one-dimensional
box
and the interparticle interaction is
repulsive.
First, we
examine
the parametric evolution of stationary states along
the cycle, which suggests that the population inversion is indeed
possible if the interparticle interaction is weak.
Also, it
is
shown that the population inversion breaks down when
the interparticle interaction is strong enough.
Second, we
numerically integrate
the time evolution equation to
confirm the picture obtained through the parametric evolution of the
stationary states.
We
provide
a theoretical explanation based on a perturbation theory to
these observations in the next section.

We assume that the Bose particles are described by the time-dependent
one-dimensional Gross-Pitaevskii equation
\begin{equation}
  \label{eq:defGP0}
  i\pdfrac{}{t}\Psi(x,t)
  = -\pdfrac{^2}{x^2}\Psi(x,t) + g|\Psi(x,t)|^2\Psi(x,t)
  ,
\end{equation}
where $\hbar=1$, $M=1/2$ and $L=2\pi$ are assumed, and $g \ge 0$
represents the strength of the effective interparticle interaction.
We impose the boundary condition $\Psi(0,t)=\Psi(L,t)=0$ and the
normalization condition $\int_0^L|\Psi(x,t)|^2dx=1$.  Let $E_n(g)$ and
$\Psi_n(x; g)$ ($n=1,2,\ldots$) denote the $n$-th eigenenergy
(chemical potential) and the corresponding stationary state for
Eq.~\eqref{eq:defGP0}.  We suppose that the system is initially in
$n$-th stationary state $\Psi_n(x; g)$.
During the cycle,
we impose
the sharp
impurity
potential (Eq.~\eqref{eq:defVi})
to Eq.~\eqref{eq:defGP0}
and vary
its
strength
$v$
slowly except at the quench point.

\begin{figure}[hbt]
  \centering
  \includegraphics[width=4.1cm]{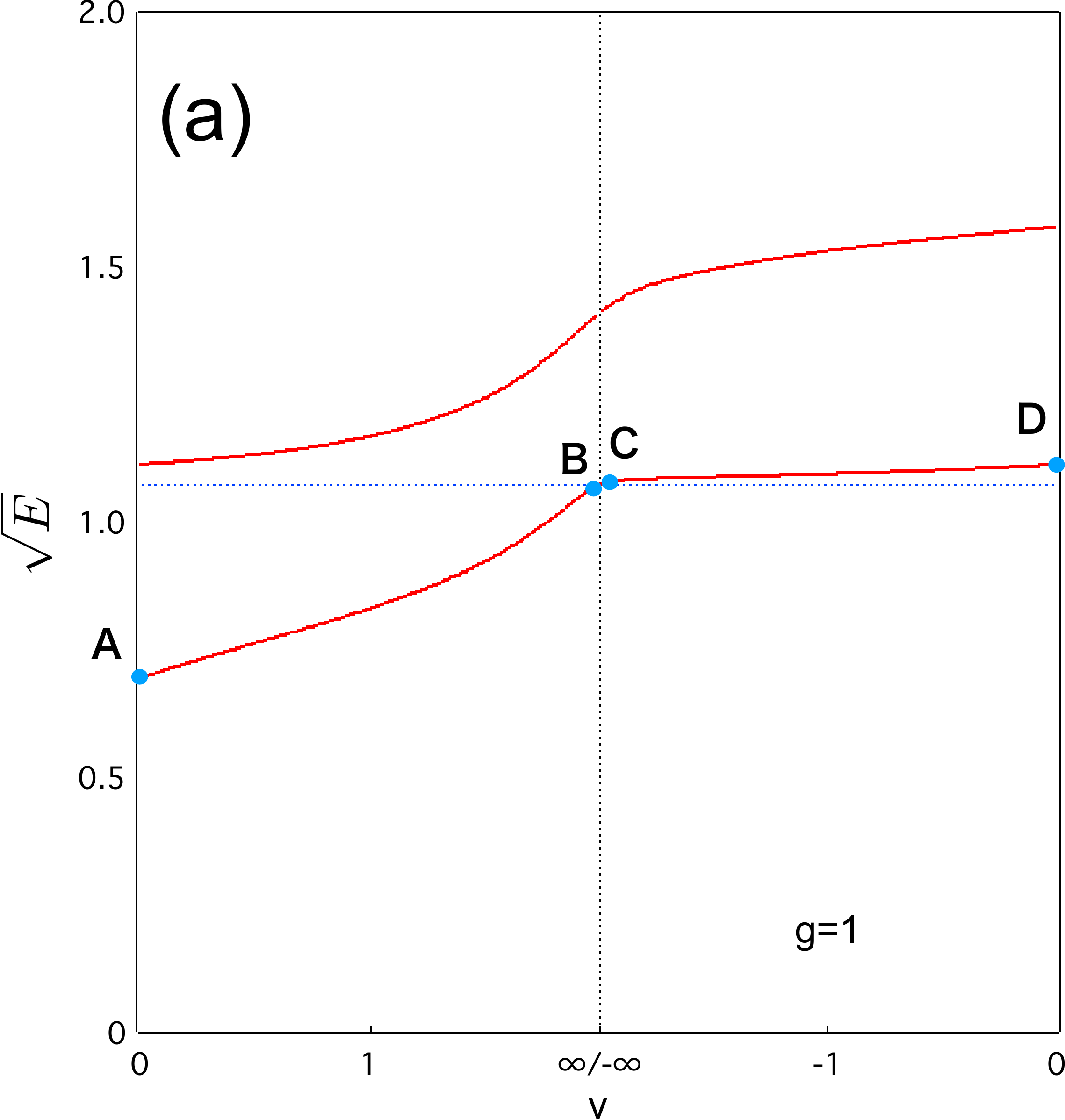}\hfill%
  \includegraphics[width=4.1cm]{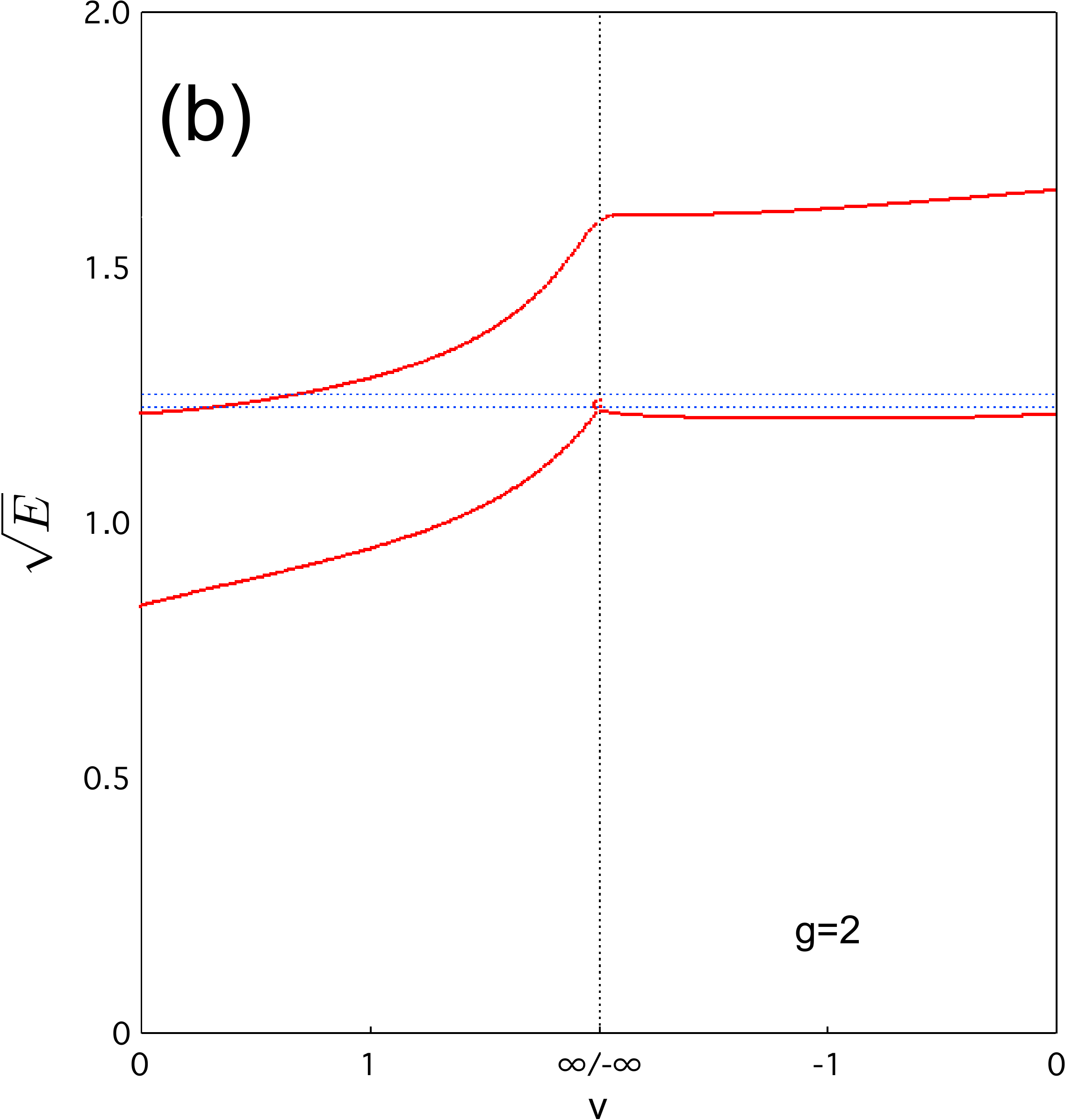}\\
  \includegraphics[width=4.1cm]{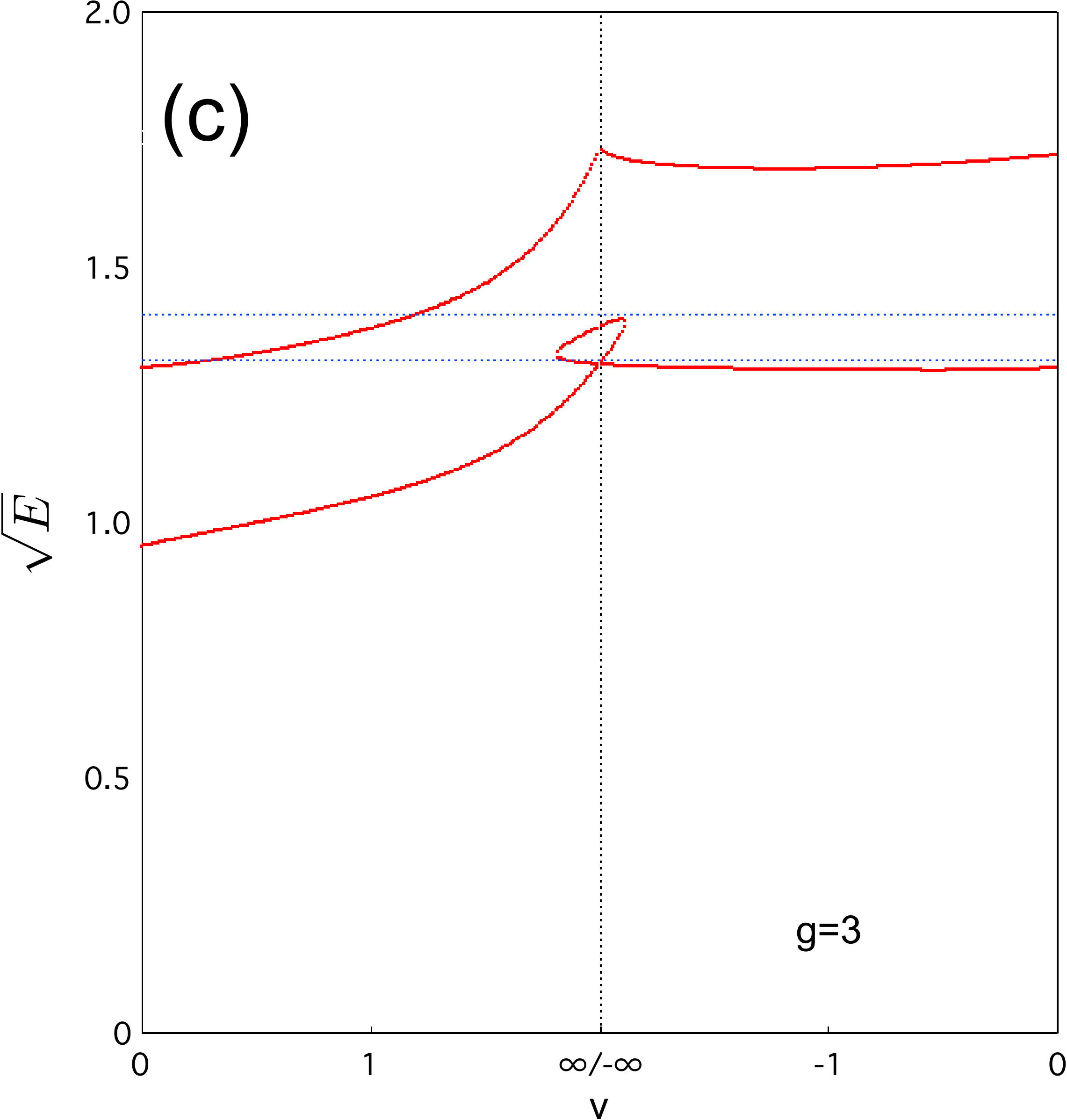}\hfill%
  \includegraphics[width=4.1cm]{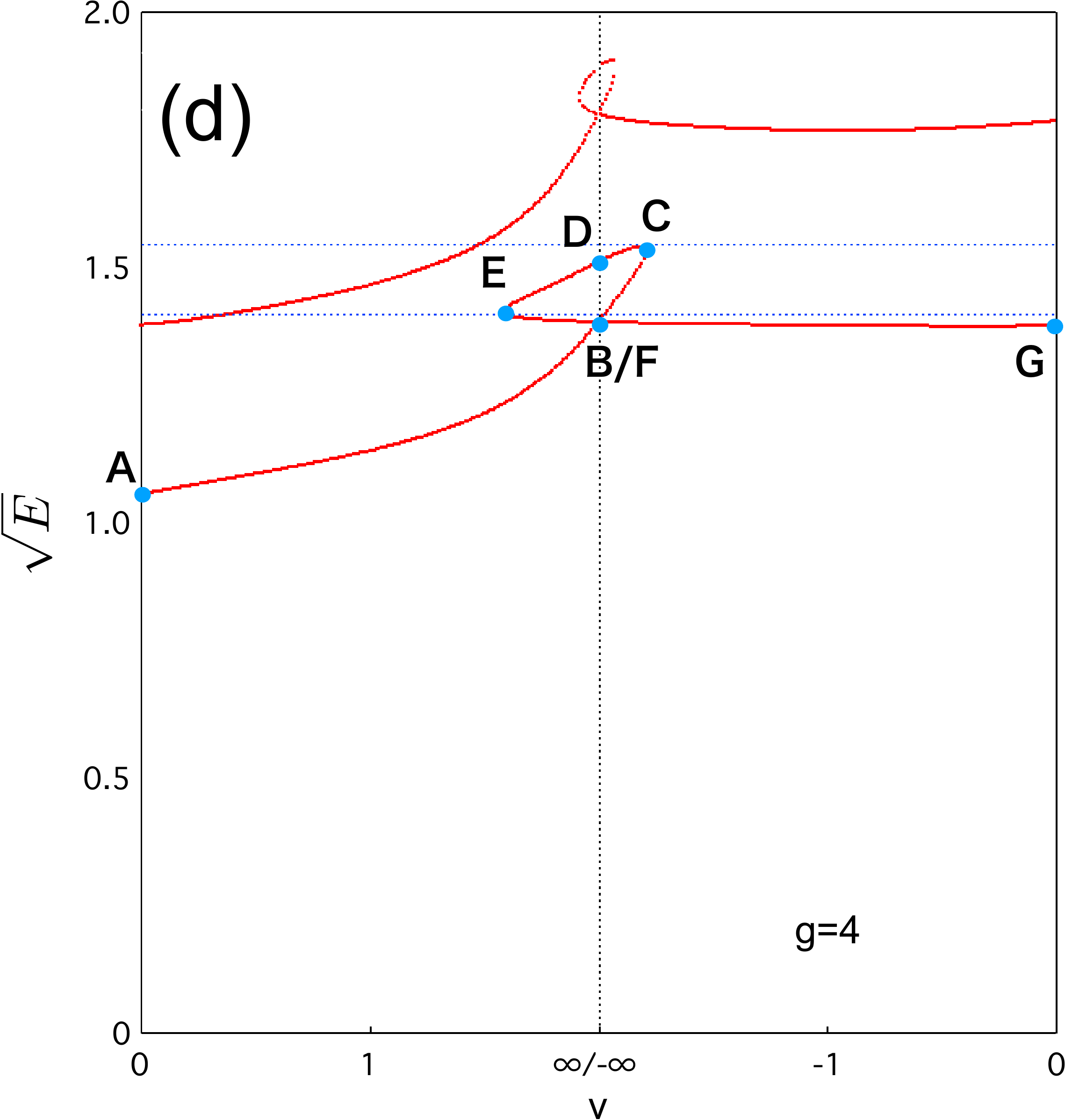}
  \caption{
    (Color online)
    Parametric evolution of the ground and first excited eigenenergies
    (chemical potentials) along $C(X=0.42L)$.
    The horizontal axis is linear in $\tan^{-1} v$.
    (a) The eigenenergies form smooth and monotonic curves at $g=1$,
    which are similar to the case $g=0$.
    (b) A tiny loop around $|v|=\infty$ appears in the ground energy curve at $g=2$.
    (c) The loop of the ground energy grows at $g=3$, while the derivative of the first excited energy seems to be discontinuous at $|v|=\infty$.
    (d) There are two noticeable loops at $g=4$.
    Blue dotted lines are the eigenenergies estimated by the two mode approximation (see, Sec.~4) of the ground branch at $|v|=\infty$.
  }
  \label{fig:nCyE}
\end{figure}

\begin{figure}[htb]
  \centering
  \includegraphics[%
  	width=4.3cm%
        ]{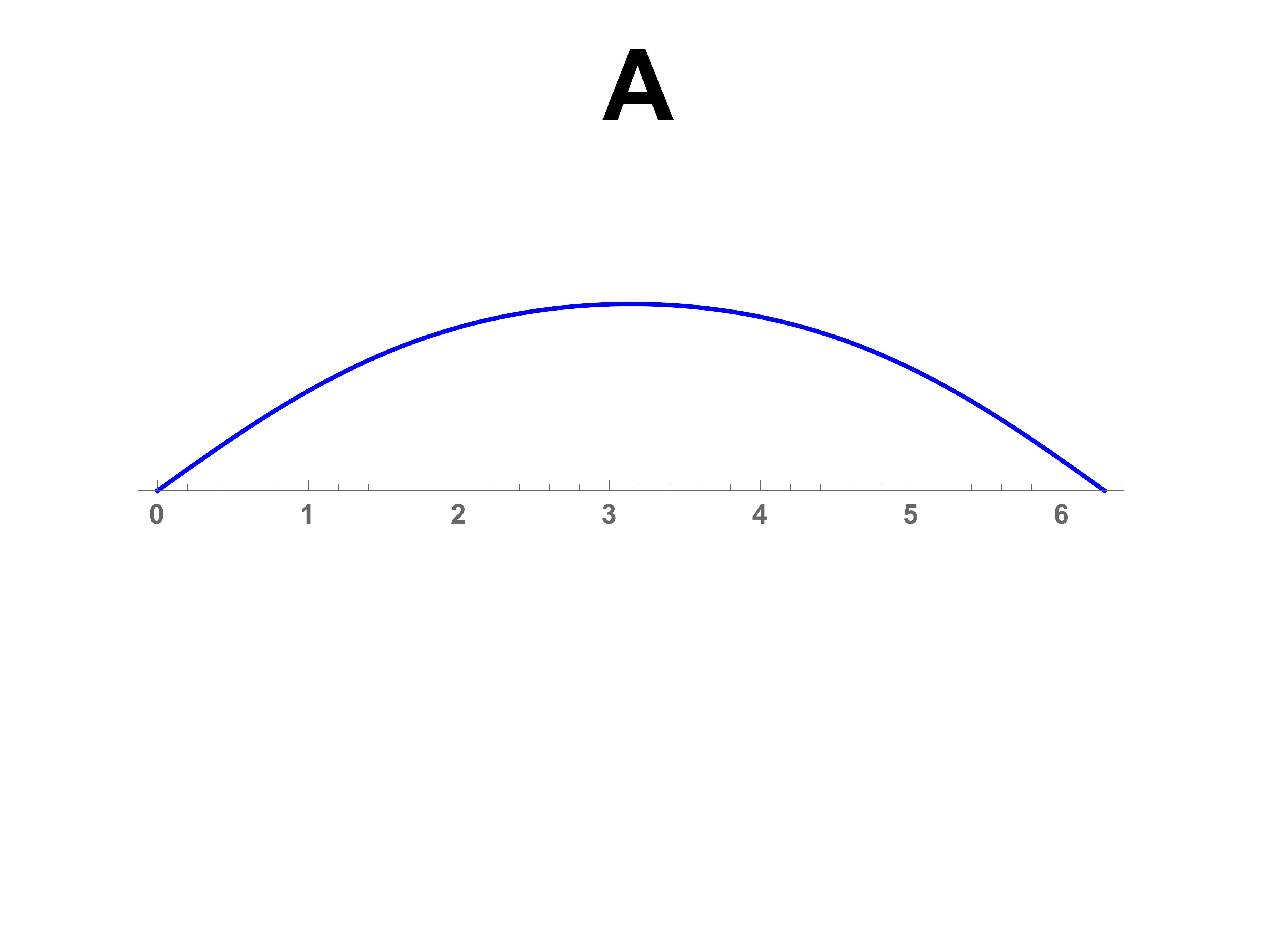}%
  \hfill
  \includegraphics[%
  	width=4.3cm%
        ]{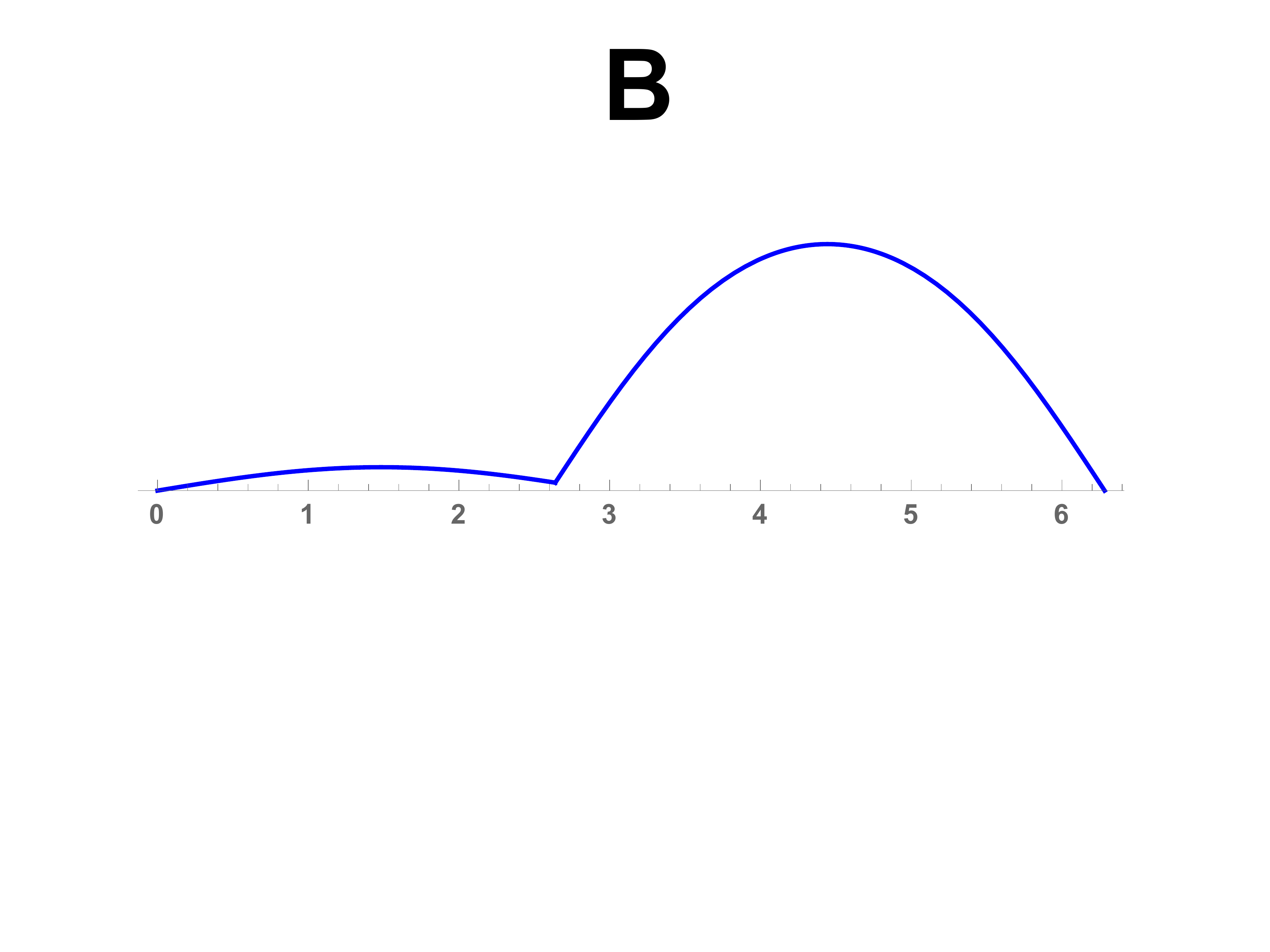}

  \includegraphics[%
   	width=4.3cm%
        ]{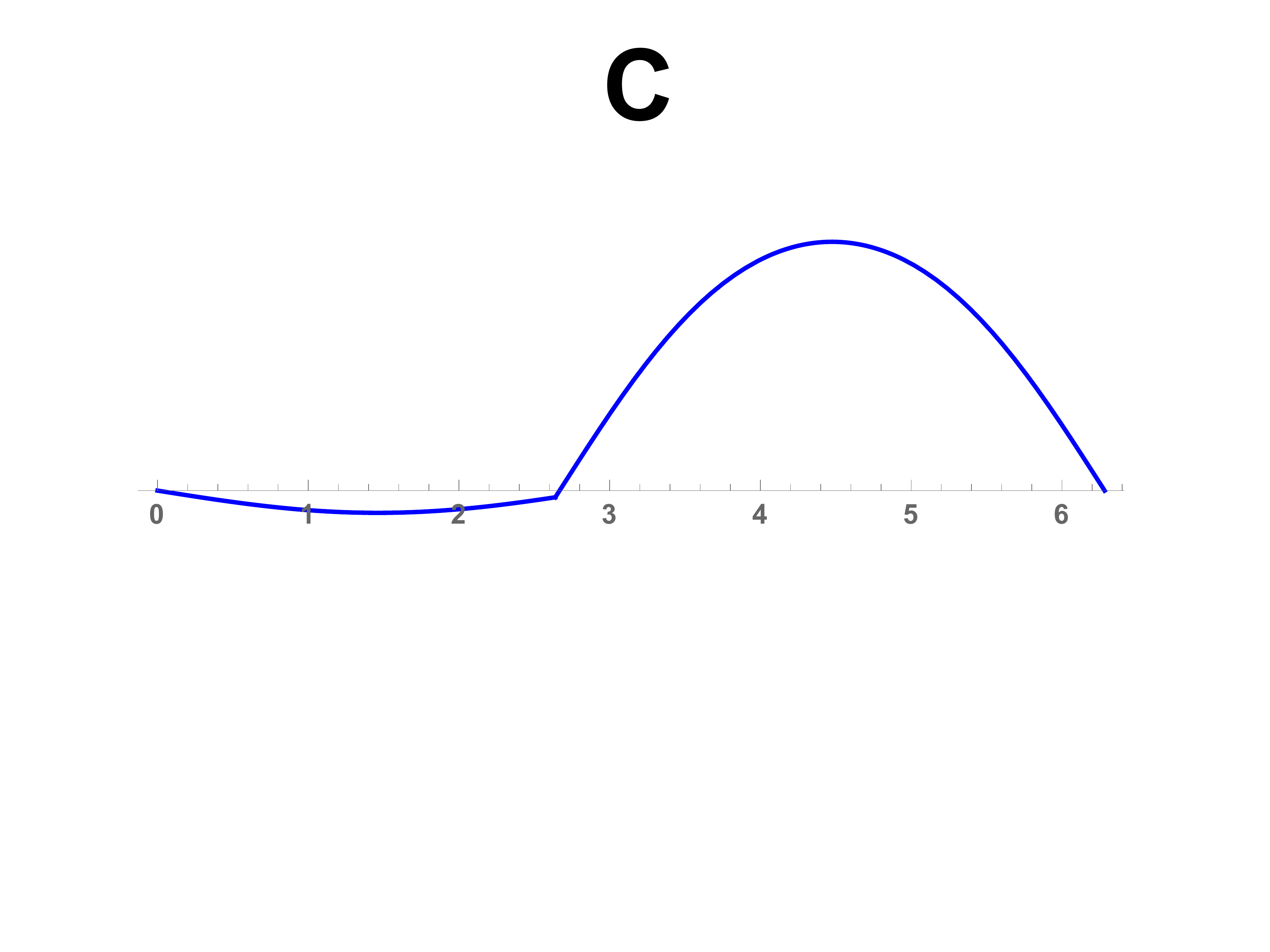}%
  \hfill
  \includegraphics[%
   	width=4.3cm%
        ]{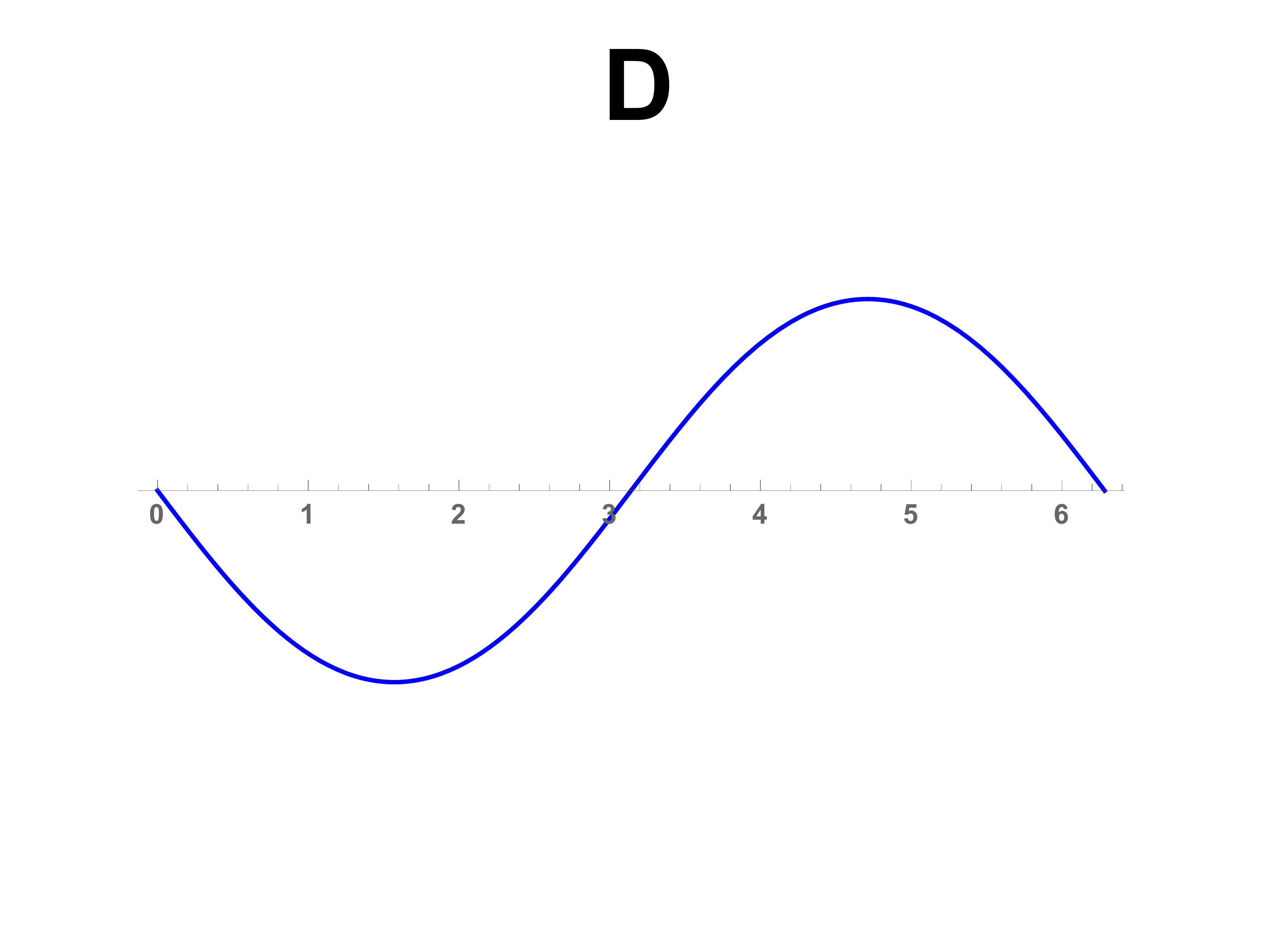}
  \caption{
    (Color online)
    Parametric evolution of stationary state $\Psi(x)$ of
    weakly interacting Bose particles along the cycle $C(X = 0.42L)$ and
    $g=1$.
    The corresponding points in $(g, E)$-plane is shown in
    Fig.~\ref{fig:nCyE}~(a).
    (A) $\Psi(x)$ is the ground state initially (i.e., $v=0$);
    (B) and (C) correspond to the case immediately before and after the quench.
    (D) the final state is the first excited state.
    The phase of $\Psi(x)$ is chosen so that
    $\Psi(x)$ is positive at right hand side of the
    sharp impurity.
  }
  \label{fig:nCyPsi_weak}
\end{figure}

\begin{figure}[hbt]
  \centering
  \includegraphics[%
  	width=2.5cm
        ]{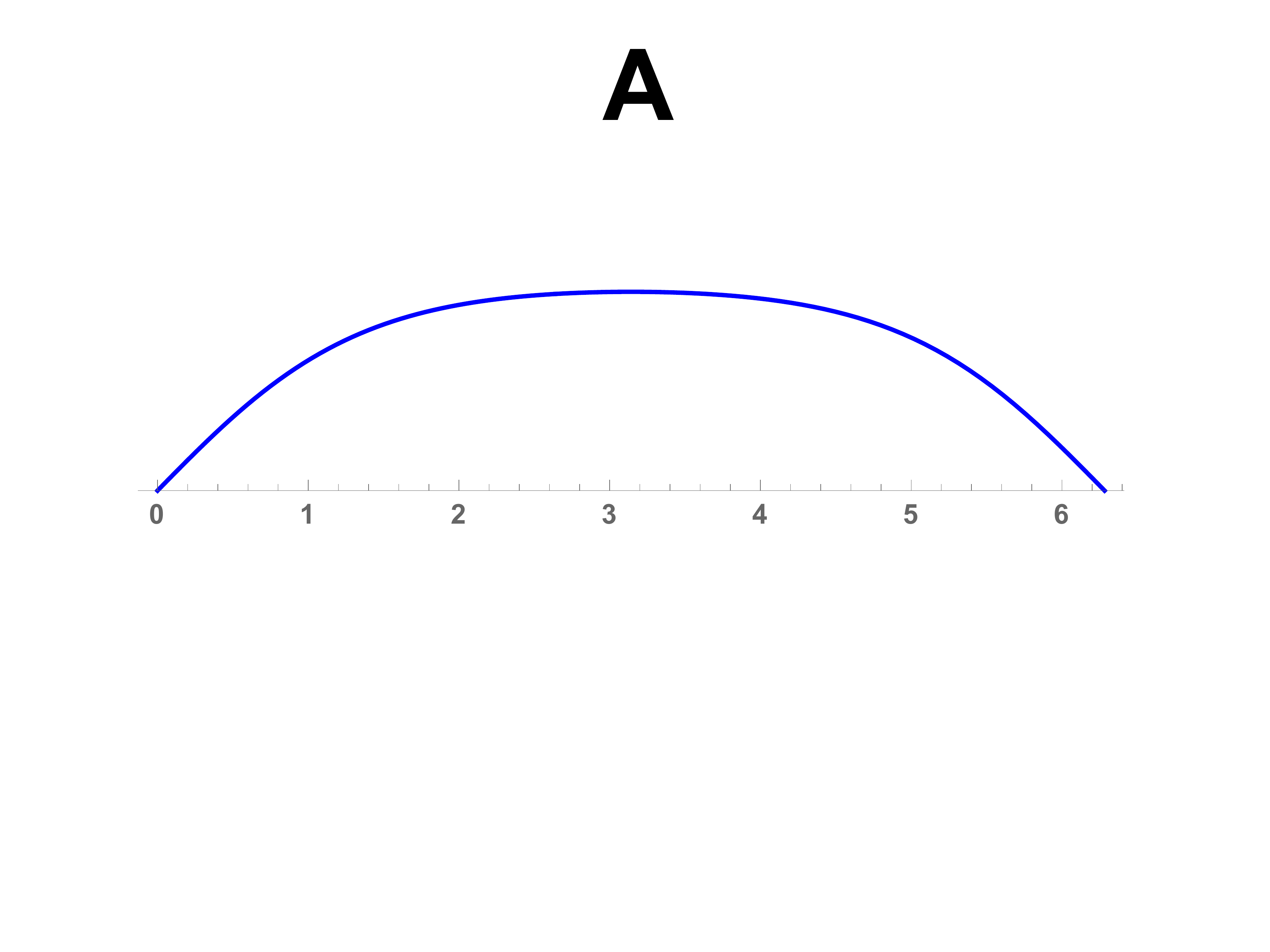}
  \includegraphics[%
  	width=2.5cm
        ]{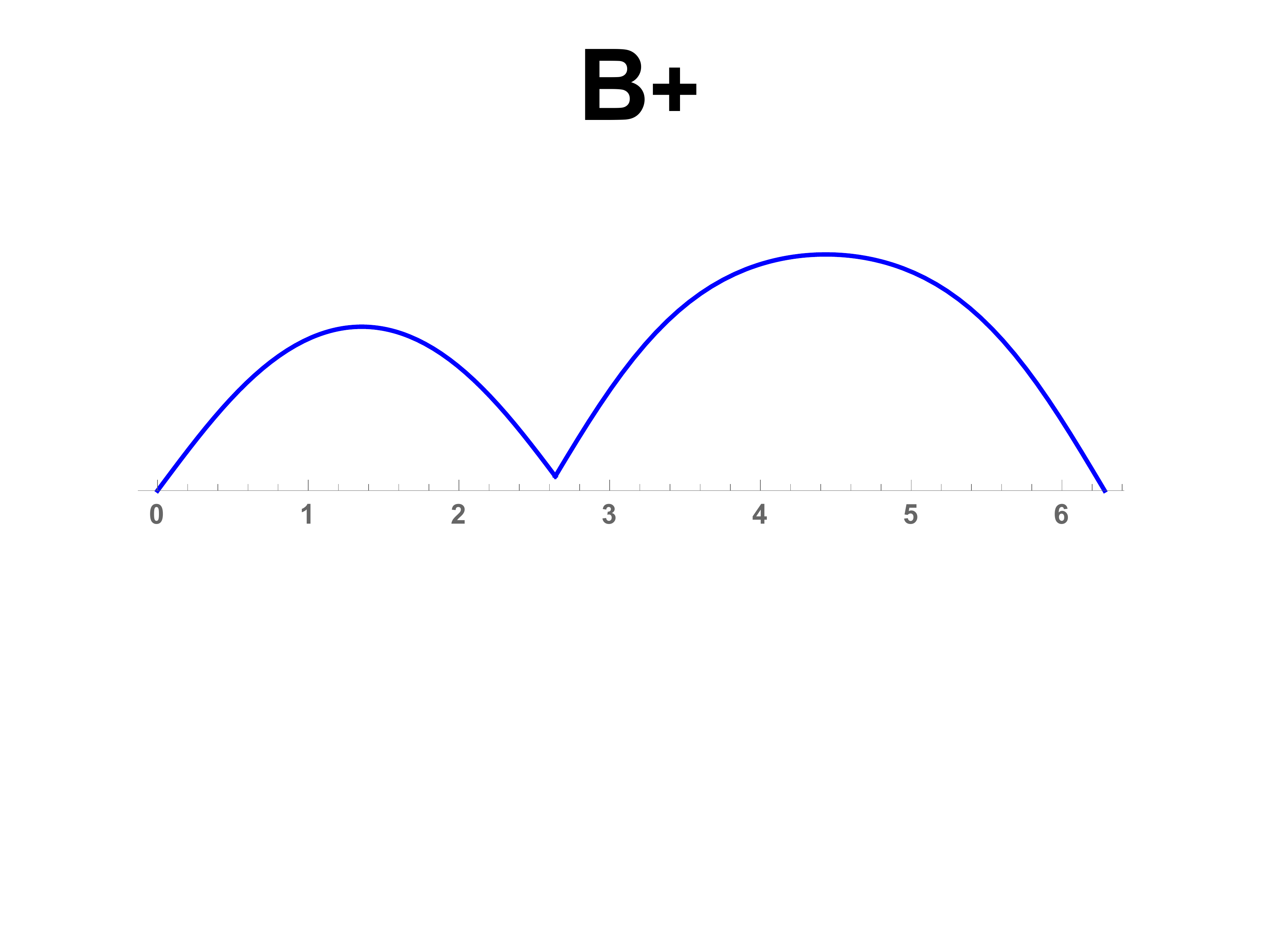}
  \includegraphics[%
  	width=2.5cm
        ]{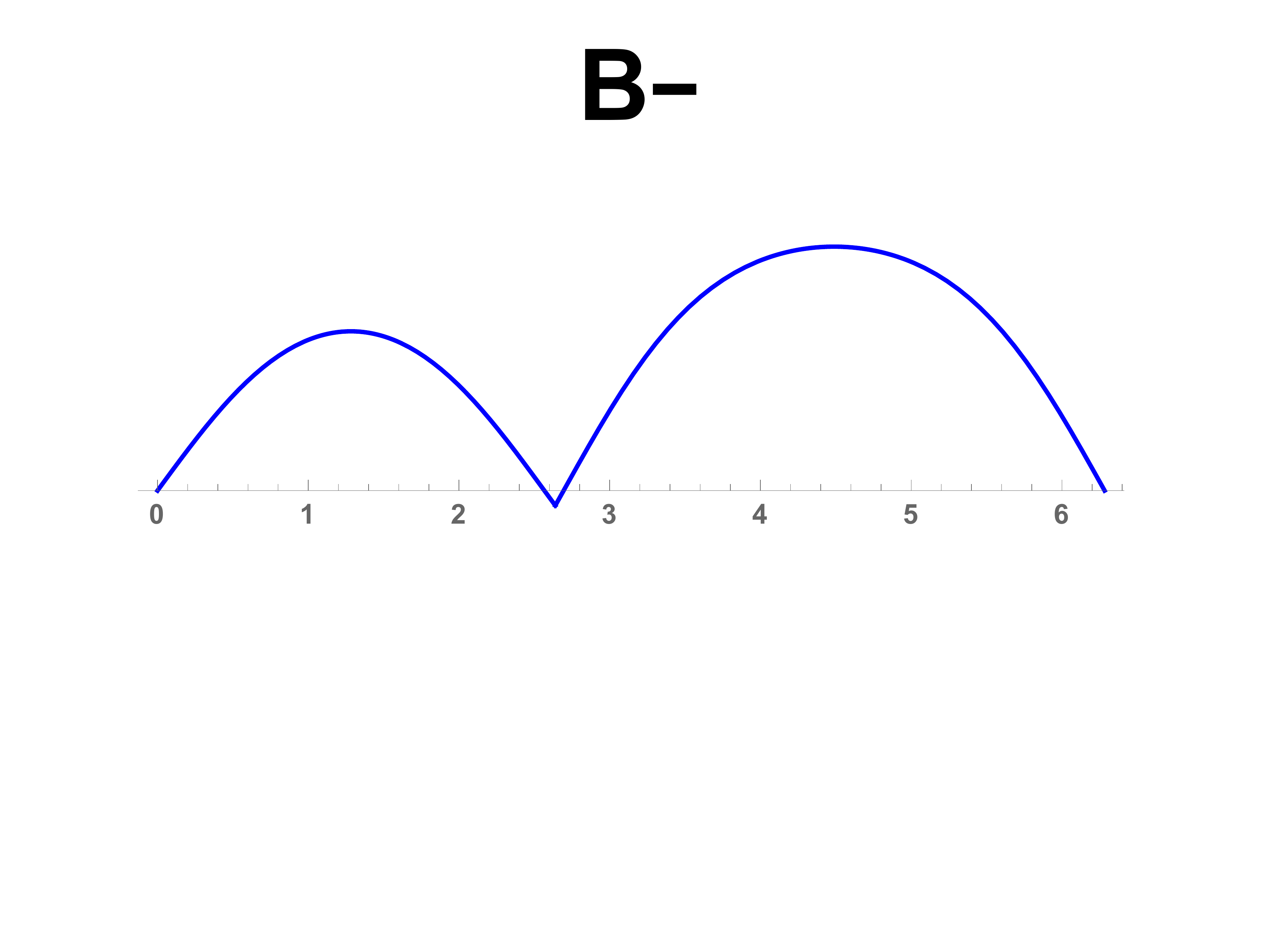}

  \includegraphics[%
  	width=2.5cm
        ]{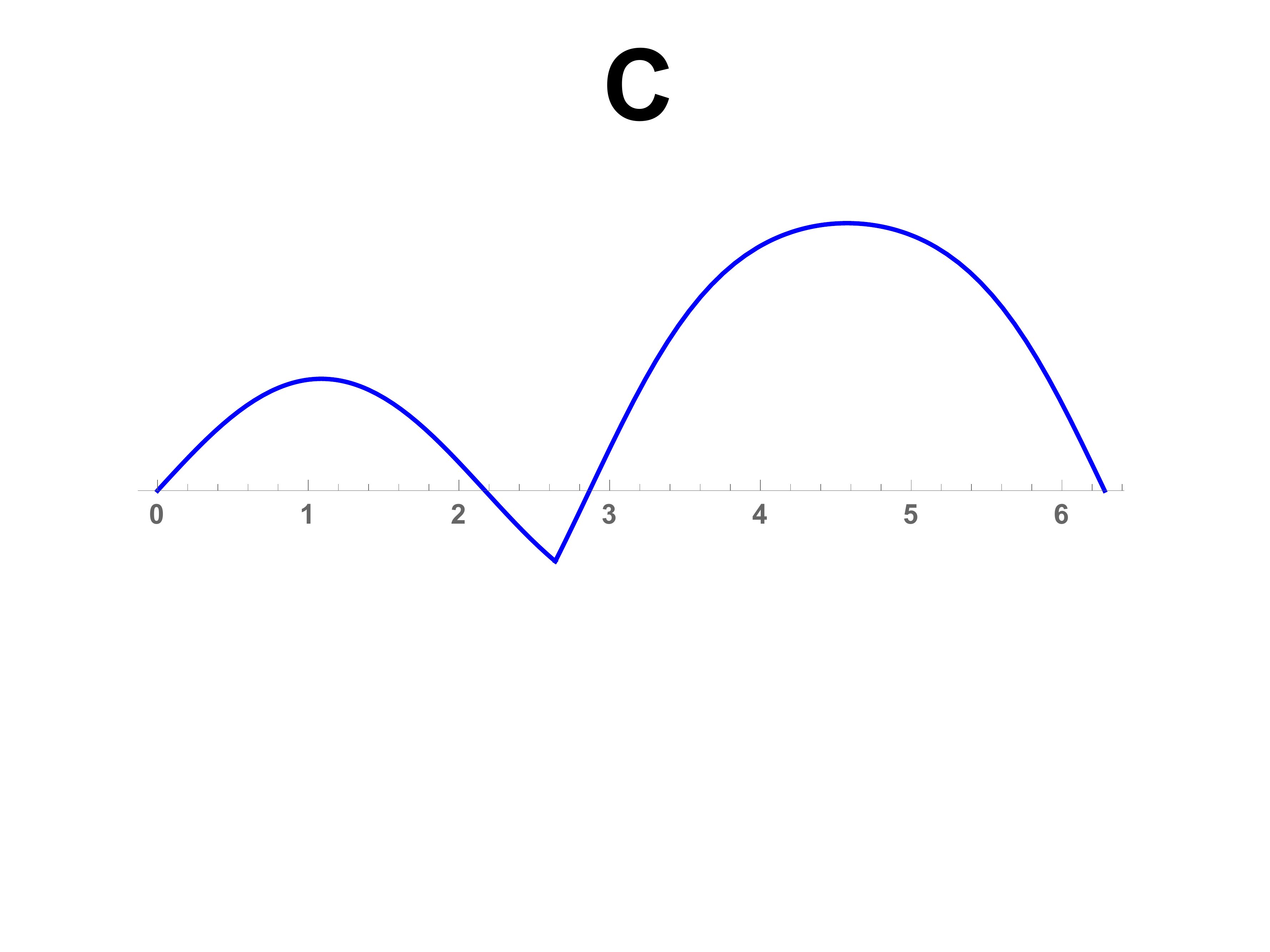}
  \includegraphics[%
  	width=2.5cm
        ]{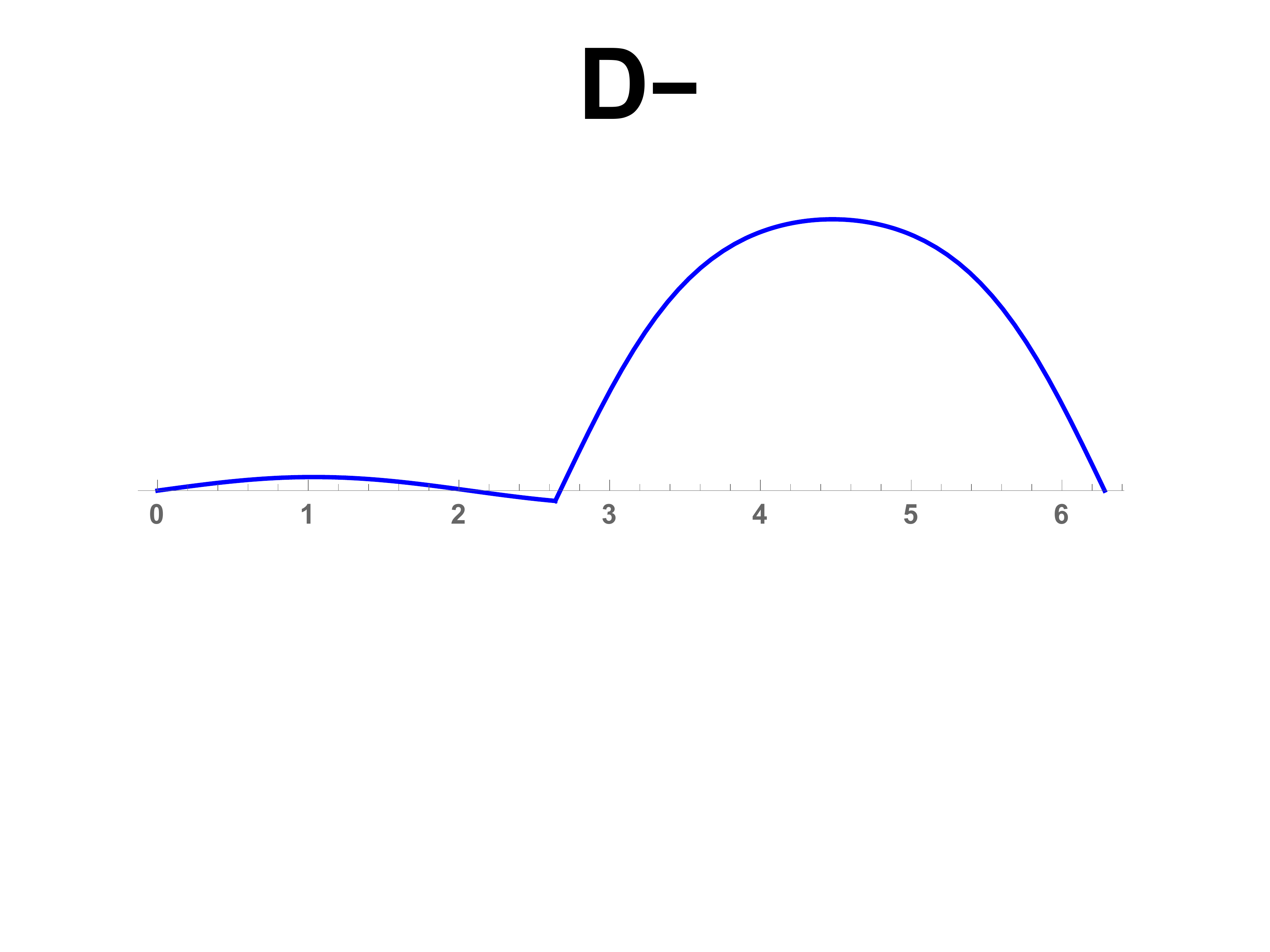}
  \includegraphics[%
  	width=2.5cm
        ]{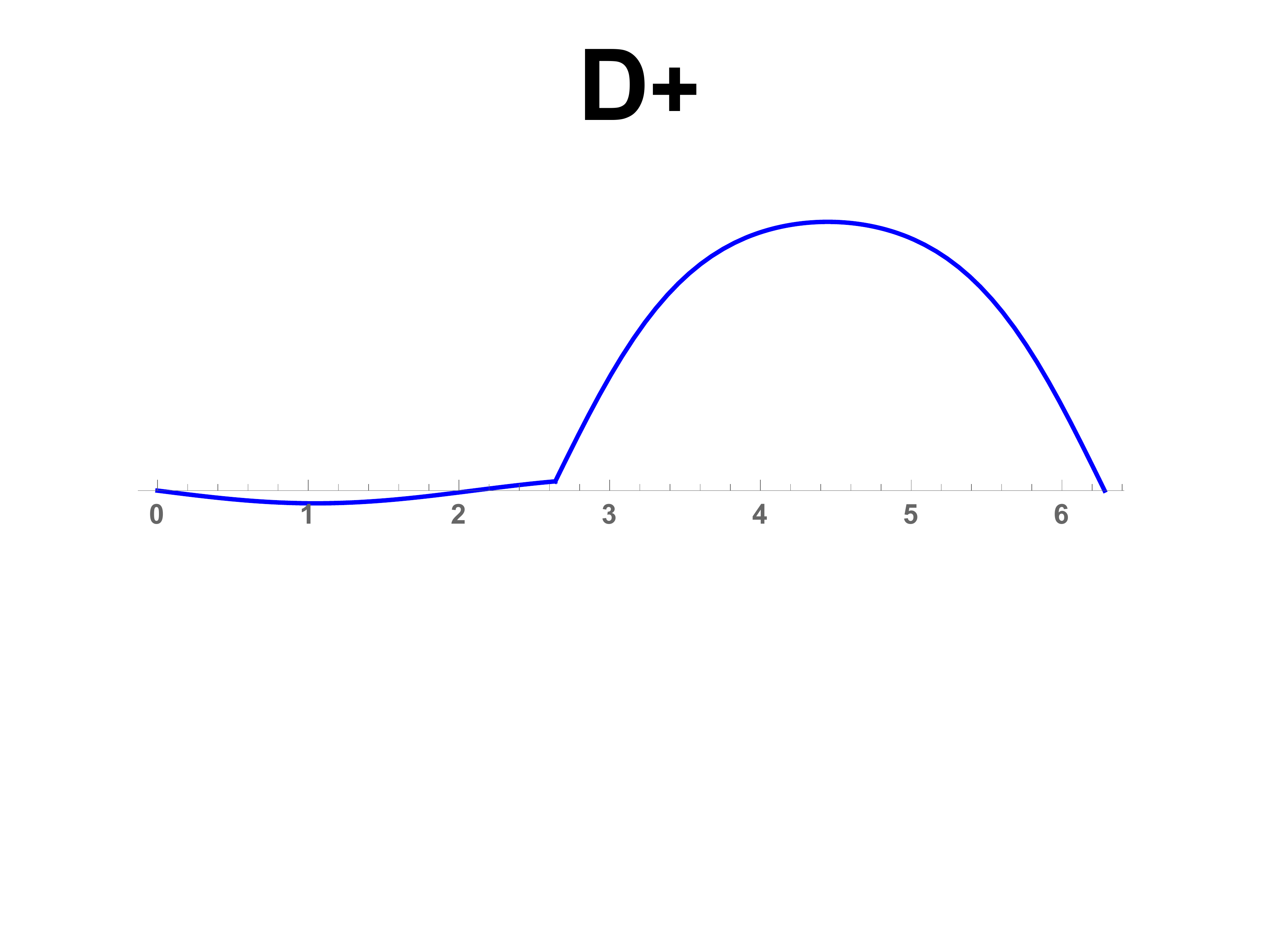}

  \includegraphics[%
  	width=2.5cm
        ]{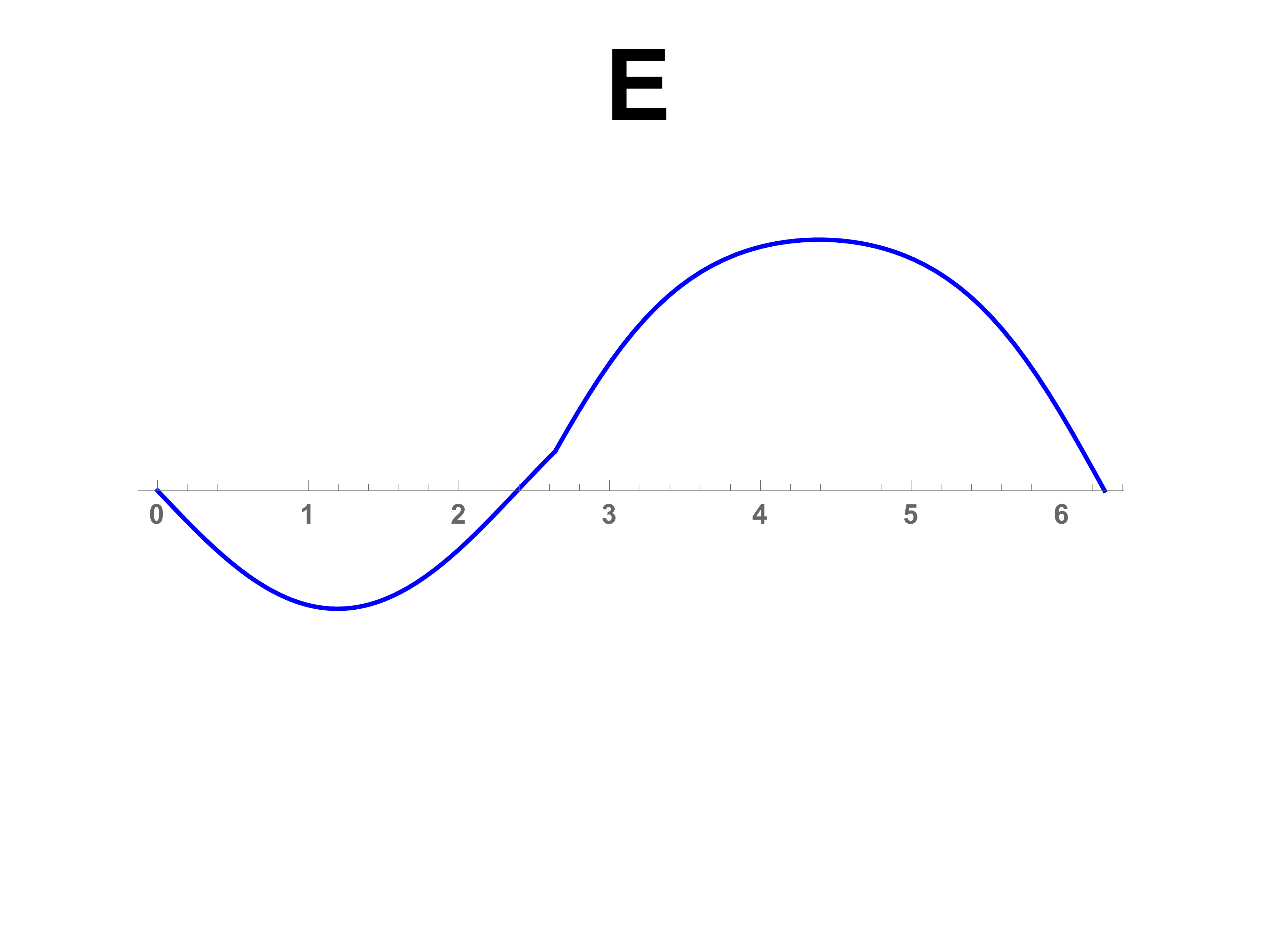}
  \includegraphics[%
  	width=2.5cm
        ]{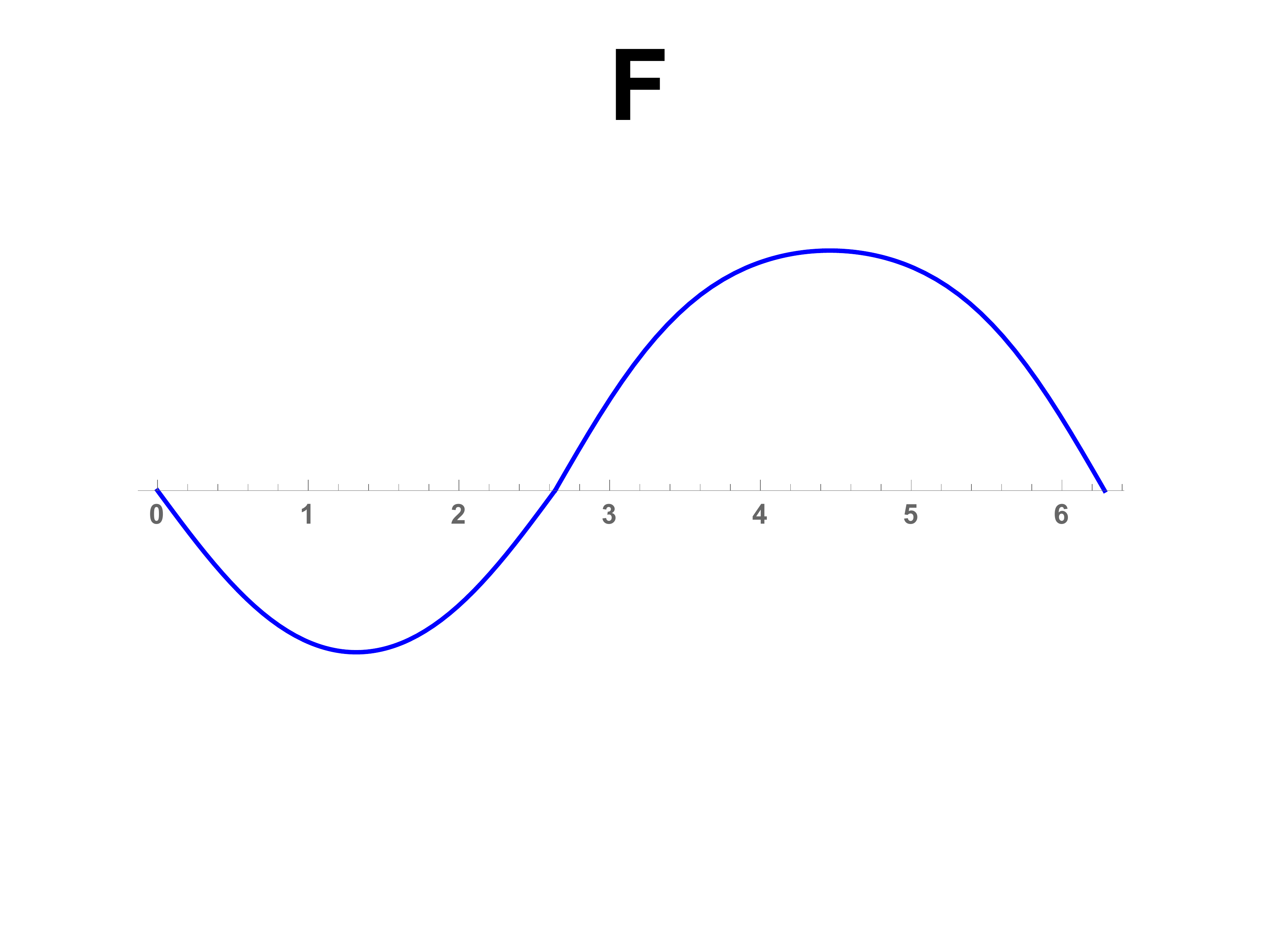}
  \includegraphics[%
  	width=2.5cm
        ]{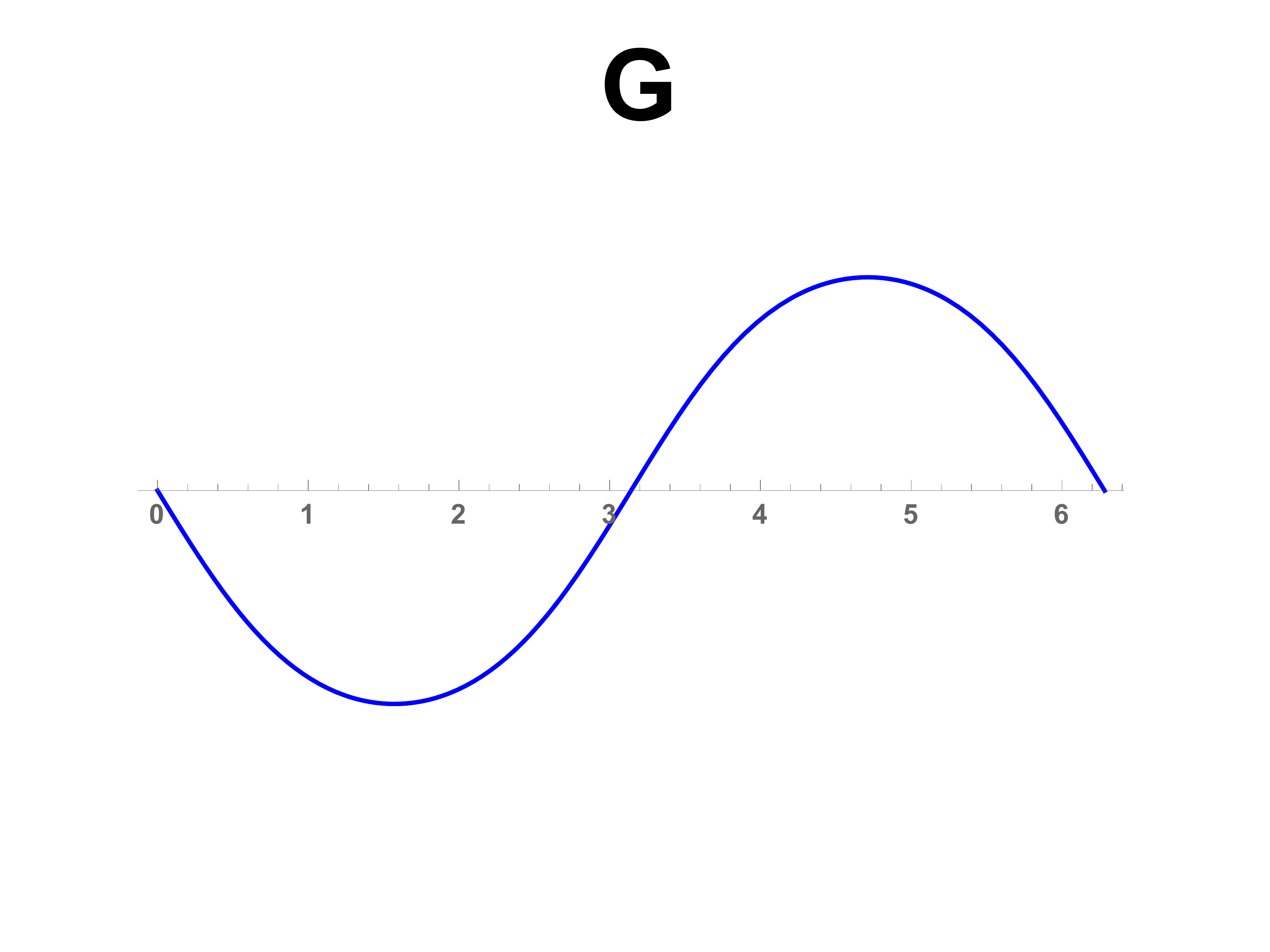}
  \caption{
    (Color online) Parametric evolution of stationary state $\Psi(x)$
    along an eigenenergy loop associated with the cycle $C(X = 0.42L)$ at
    $g=4$, which correspond to the case shown in Fig.~\ref{fig:nCyE}~(d).
    The loop connects the initial point A and the final point G of the
    cycle $C(X)$. Around the quench points $B$ and $D$, the sign of $v$ is
    indicated by the suffix $\pm$.
  }
  \label{fig:nCyPsi}
\end{figure}

To find the stationary states of the system where the position $X$ and
strength $v$ of the
impurity
potential~(Eq.~\eqref{eq:defVi}) are ``frozen'',
we examine the time-independent Gross-Pitaevskii equation with
the
impurity
potential
\begin{equation}
  \label{eq:tiGP}
  \left(-\pdfrac{^2}{x^2} + g|\Psi(x)|^2
    + V_{\rm i}(x; v)\right)\Psi(x)
  = E\Psi(x)
  .
\end{equation}
Examples of parametric evolution of $n$-th
eigenenergy
along the cycle
$C(X)$ are shown in Fig.~\ref{fig:nCyE}. Corresponding parametric
evolution of stationary states are shown in
Figs.~\ref{fig:nCyPsi_weak} and~\ref{fig:nCyPsi}.

When the interparticle interaction $g$ is small enough, the quasi-adiabatic
cyclic operation induces
the population inversion.
The parametric evolution of eigenenergy along the cycle connects the
initial eigenenergy $E_{1}(g)$ with $E_{2}(g)$ (see,
Fig.~\ref{fig:nCyE}~(a)). The connection is equivalent to the case of
noninteracting particles (see, Fig.~\ref{fig:linearCq}).  This allows
us to infer the parametric evolution of eigenfunction whose initial
condition is the ground state $\Psi_1(x; g)$ of Eq.~\eqref{eq:defGP0},
i.e., at $g=0$ (see, Fig.~\ref{fig:nCyPsi_weak} (A)).
While the stationary state is nodeless during
the strength of the sharp
impurity potential
$v$ is positive and finite,
the state becomes localized at a side of the
impurity
as $v$ become
larger (Fig.~\ref{fig:nCyPsi_weak} (B)). Immediately before the
quench,
i.e, $v=\infty$,
the localization completes~\cite{GeaBanacloche-AJP-70-307}.
The state is
unchanged during
%
the flip of the potential strength from
$v=\infty$ to
$v=-\infty$~\cite{Kasumie-PRA-93-042105}. As $v$ is
slightly increased from $-\infty$, the stationary state extends to the
other side of the
impurity
to produce a node (Fig.~\ref{fig:nCyPsi_weak}
(C)). The resultant stationary state has a single node while $v$ is finite
(Fig.~\ref{fig:nCyPsi_weak} (D)). This is the reason why the
destination of the stationary state at the end of the cycle is the
first excited state $\Psi_2(x;g)$.

On the other hand, when the interparticle interaction is large,
discrepancies from the linear case become significant. In particular,
as is seen in Figs.\ref{fig:nCyE} (c) and (d),
the parametric evolution of a stationary energy involves a loop structure
that emanates from the quench point
$|v|=\infty$
in $C(X)$.
We note that
loop structures are
often observed in the studies of time-independent Gross-Pitaevskii
equation~\cite{Wu-PRA-61-023402,Damski-PRA-65-013604,Liu-PRA-66-023404,Mueller-PRA-66-063603,Kunimi-PRA-91-053608}.

The corresponding parametric evolution of the stationary state along
the loop is explained in Fig.~\ref{fig:nCyPsi}.
In the vicinity of the quench, the wavefunction extends to both sides of
the sharp impurity potential,
which is a distinctive feature of the case with stronger
interparticle interaction (${\rm B}_{+}$).  Across the
quench point
the wavefunction
smoothly connect to the lower branch of the loop to acquire two nodes
(${\rm B}_{-}$). As $v$ increased, the lower branch arrives the
extremum point (C) to connect the uppermost branch, where the
wavefunction localizes at a side of the
impurity potential,
where the stationary
state mimics the one in the linear system. As $v$ decreased to follow the
uppermost branch, the stationary state become localized at a side of
the impurity
(${\rm D}_{-}$). At the quench point in the uppermost branch, the
localization completes. Across the quench, the number of the nodes of
the stationary decreases
from $2$ (${\rm D}_{-}$) to $1$ (${\rm D}_{+}$).
Then the uppermost branch arrives another extremum
point (E), where the state delocalize again to connect the final
branch at (F), which smoothly connects the first excited state $\Psi_2(x;g)$,
see (G).

We expect that the loop structure disturbs
the adiabatic
evolution, as reported in Refs.~\cite{Wu-PRA-61-023402,Liu-PRA-66-023404},
and thus hinders the population inversion. This is because the
stationary state is transformed into a non-stationary state when
the adiabatic time evolution has to departs from the parametric evolution
by having a loop.

To clarify whether the adiabatic time evolution along $C(X)$
really occurs, we
numerically
examine the linear stability of the stationary
states in $C(X)$ by diagonalizing
the Bogoliubov equation~\cite{Bogolubov-JPhysUSSR-11-23,Fetter-AP-70-67}
corresponding to the stationary solutions.
%
Also, an analytical study on the linear stability for the quench point
is shown in Appendix~\ref{sec:Bogoliubov}.

%
When the interparticle interaction is small (see, Figs.~\ref{fig:nCyE}~(a)
and \ref{fig:nCyPsi_weak}), we find that the stationary states are linearly
stable along $C(X)$. Hence we may expect that the adiabatic time evolution
remains intact for the weakly interacting case.

Meanwhile, when the interparticle interaction is larger to form
the loop structure as shown in
Figs.~\ref{fig:nCyE}~(d) and \ref{fig:nCyPsi},
we find that
the stationary state is linearly stable within the intervals
from $v=0$ through $v=\pm\infty$, i.e., from ${\rm A}$ to ${\rm B}$
and from ${\rm G}$ to ${\rm F}$ in Fig.~\ref{fig:nCyE}~(d),
and the uppermost branch of the loop
(from ${\rm C}$ to ${\rm E}$).
On the other hand, the stationary state is linearly
unstable at the lower part of the loop
(i.e, from ${\rm B}$ to ${\rm C}$ and ${\rm E}$ to ${\rm F}$).

The result of the linear stability analysis for stronger interparticle
interaction suggests that
the adiabatic time evolution whose initial condition is
the ground state $\Psi_1$ remains intact until the quench point,
i.e., within the interval ${\rm A}$ to ${\rm B}$.
After the system passes the quench point,
the adiabatic time evolution breaks down during
the interval ${\rm B}$ to ${\rm C}$ due to the linear instability.
We note that the emergence of the unstable stationary state at the lower
branch of the loop structure in the Brillouin zone is reported in
Refs.~\cite{Wu-NJP-5-104,Danshita-PRA-75-033612}.
We
also
note that this is a distinctive point from
the instabilities in
the conventional studies~\cite{Wu-PRA-61-023402,Liu-PRA-66-023404}
of adiabatic time evolution along the loop
structure, where the linearly unstable region appears only at the uppermost
of the loop structure.
After the point ${\rm C}$, the adiabatic time evolution is impossible since
the stationary solution cannot be adiabatically extended
anymore~\cite{Wu-PRA-61-023402,Liu-PRA-66-023404}.

\begin{figure}[htb]
  \centering
  \includegraphics[%
	width=8.66cm
        ]{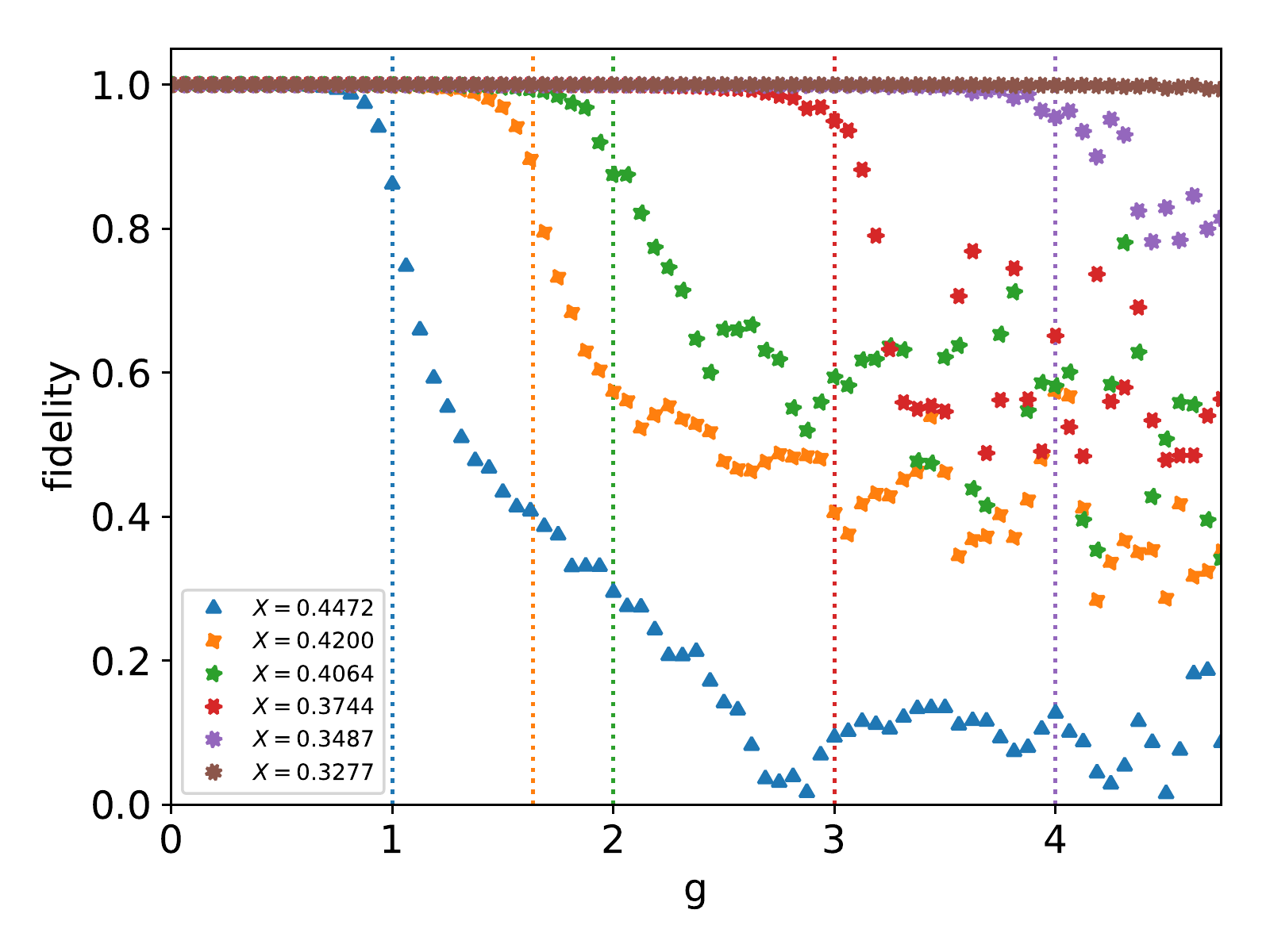}
  \caption{
    (Color online)
    Population inversion probability from the ground state through
    $C(X)$.
    Vertical dashed lines indicate the critical point predicted by the
    two-mode approximation (see, Sec.~4).
  }
  \label{fig:pop2}
\end{figure}

We test the scenario above through numerical integration of the
time-dependent Gross-Pitaevskii equation along the almost-adiabatic
cycle $C(X)$.
We show our numerical result for various values of $X$ in
Fig~\ref{fig:pop2}, where the initial condition is prepared to be in
the ground state $\Psi_1(x;g)$ of Eq.~\eqref{eq:defGP0}. We
numerically evaluate the fidelity for the population inversion
$|\bracket{\Psi_2(g)}{\Psi}|^2$, where $\Psi$ is a state after the
completion of the cycle. Since the finial states may not be
stationary, we depict the time-average of the fidelity probability
after the completion of the cycles.

From Fig~\ref{fig:pop2}, we conclude that the population inversion
fails if the value of $g$ exceeds a critical value $g_{\rm c}$, which
depends on the position of the
impurity
potential $X$. Moreover, when
we restrict the case $0 < X < L/2$, $g_{\rm c}$ becomes larger as $X$
become smaller.

We make a remark on the integration of time-dependent Gross-Pitaevskii
equation along $C(X)$, where we introduce an approximation
for the quench of
impurity potential.
We keep $v$, the strength of the
impurity potential.
finite.
Namely, $v$ is increased from $0$ to $v_{\rm max}$ with a finite
velocity $dv/dt$ during the first process $C_{\rm s}(0, v_{\rm max})$. At the
quench, $v$ is suddenly changed from
$v_{\rm max}$ to $-v_{\rm max}$. Then, during $C_{\rm s}(-v_{\rm max}, 0)$, the
value of $v$ is increased from $-v_{\rm max}$ to $0$ with non-zero
velocity $dv/dt$.
Although this induces a tiny nonadiabatic error during the quench, as
is seen from Fig.~\ref{fig:pop2}, the error is far less important
than the nonlinear effect.

\section{Two-mode approximation
  at the
  quench
  point}
We discuss our numerical results
in the previous section
with an approximate
theory. In particular, we would like to clarify the reason why the
population inversion breaks down as the strength of the interparticle
interaction becomes stronger (Fig.~\ref{fig:pop2}).
A key ingredient
must be the emergence of the loop structure in
$(g,E)$-plane (Fig.~\ref{fig:nCyE}). To identify the loop structure,
we examine the quench point
$|v|=\infty$, because a loop emanates from a point at $|v|=\infty$ in
$(g,E)$-plane. This analysis allows us to infer the loop structure, as
long as the loop is small enough.

In the following,
we utilize a two-mode approximation. Namely, we assume that the
stationary wavefunction $\Psi(x)$ is a superposition of two
eigenfunctions $\psi_j(x)$ ($j=0,1$) of the noninteracting
system. Since the
infinitely strong impurity
divide the box completely~\cite{GeaBanacloche-AJP-70-307}, as
suggested in Fig.~\ref{fig:nCyE}, we utilize the unperturbed
eigenfunction $\psi_j(x)$ that is localized at the left or right side
of the
impurity.

For example, to examine the stationary states that are associated with
the ground state at the initial point of the cycle,
we assume that $\psi_0(x)$ and $\psi_1(x)$
describe the ground state of a single particle confined within the right
and left boxes, respectively, i.e,
\begin{align}
  \psi_0(x)
  &=
    \begin{cases}
      0 & \text{for $0 \le x \le X$}
      \\
      \sqrt{\frac{2}{L-X}} \sin\frac{\pi(x-X)}{L-X}
      &\text{for $X \le x \le L$}
    \end{cases}
          ,
  \\
  \psi_1(x)
  &=
    \begin{cases}
      \sqrt{\frac{2}{X}} \sin\frac{\pi x}{X}
      &\text{for $0 \le x \le X$}
      \\
      0 & \text{for $X \le x \le L$}
    \end{cases}
          ,
\end{align}
whose eigenenergies are
$E_0=E_{\rm g}/r'^2$ and $E_1=E_{\rm g}/r^2$, respectively,
where $r=X/L$, $r'=(L-X)/L$, and $E_{\rm g}= \hbar^2\pi^2/(2ML^2)$.

In the following, we assume $0 < X < L/2$, which
implies $E_0 < E_1$, i.e., the ground state $\psi_0(x)$ under the
presence of the
infinitely strong impurity
localizes at the right side of the impurity.

From the two-mode assumption that $\Psi(x,t) =
\Psi_0(t)\psi_0(x)+\Psi_1(t)\psi_1(x)$ satisfies the time-dependent
Gross-Pitaevskii equation, we obtain the time-evolution equation for
the amplitudes $\Psi_j(t)$ ($j=0,1$):
\begin{equation}
  i\frac{d}{dt} \Psi_j(t)
  = E_j \Psi_j(t)
  + g\int_0^L\psi^*_j(x)\left|\Psi(x,t)\right|^2\Psi(x,t) dx
  .
\end{equation}
Because $\psi_0(x)$ and $\psi_1(x)$ have no overlap in the position
space, i.e., $\psi_0(x)\psi_1(x)=0$ holds, and are real, we obtain
\begin{equation}
  i\frac{d}{dt} \Psi_j
  = E_j\Psi_j + g\int_0^Ldx\left\{\psi_j(x)\right\}^4
  \left|\Psi_j\right|^2\Psi_j
  .
\end{equation}
Hence the nonlinear Scr\"odinger equation for $\Psi_j$ is
\begin{equation}
  \label{eq:nls2}
  i\frac{d}{dt} \Psi_j
  = \left(E_j + g c_j|\Psi_j|^2\right)\Psi_j
  ,
\end{equation}
where
$c_0\equiv 3g/\{2(L-X)\}$
and
$c_1\equiv 3g/(2X)$.
We also impose the normalization condition
$|\Psi_0|^2+|\Psi_1|^2=1$.

The stationary solutions of Eq.~\eqref{eq:nls2} are classified into
two groups.  First, there are two localized solutions $(\Psi_0,
\Psi_1) = (1,0)$ and $(0,1)$, whose eigenenergies are
\begin{gather}
  \label{eq:Elocalized}
  E_0(g)
  = E_0 + g c_0
  ,\quad\text{and}\quad
  E_1(g)
  = E_1 + g c_1
  ,
\end{gather}
respectively.

Second, the other two solutions $\Psi_{\pm}$ are
\begin{equation}
  \label{eq:nls2c}
  \begin{bmatrix}
    \Psi_{\pm,0} \\ \Psi_{\pm,1}
  \end{bmatrix}
  =
  \begin{bmatrix}
    \sqrt{r' - (E_0-E_1)({2Lrr'})/(3g)}
    \\
    \pm\sqrt{r + (E_0-E_1)({2Lrr'})/(3g)}
  \end{bmatrix}
  ,
\end{equation}
which are delocalized to the both side of the impurity.
The corresponding
eigenenergies are doubly degenerate
\begin{equation}
  \label{eq:Edeloc}
  E_{\pm}(g) = r' E_0 + r E_1 + \frac{3g}{2L}
  .
\end{equation}
We explain the condition that the stationary states Eq.~\eqref{eq:nls2c} are
physical, i.e.,
$0\le |\Psi_0|^2, |\Psi_1|^2 \le 1$ holds. First, we introduce the
critical points
\begin{align}
  g_0 &= \frac{2(L-X)}{3}(E_1-E_0)\\
  g_1 &= -\frac{2X}{3}(E_1-E_0)
        ,
\end{align}
where $|\Psi_j|^2=1$ holds if $g=g_j$ ($j=0,1$).
Since we assume $X<L/2$,
the physical condition for the stationary solution Eq.~\eqref{eq:nls2c} is
summarized as $g\ge g_0$ or $g\le g_1$.
Also, as we restrict the case that the interparticle interaction is repulsive,
i.e., $g\ge 0$, the delocalized solutions
Eq.~\eqref{eq:nls2c}
exist
only when $g\ge g_0$.

From the linear stability analysis (Bogoliubov analysis)
of Eq.~\eqref{eq:nls2},
whose details are shown in Section~\ref{sec:Bogoliubov},
we find that
the localized solutions $(\Psi_0, \Psi_1) = (1,0)$ and $(0,1)$ are stable.
On the other hand, the delocalized solutions Eq.~\eqref{eq:nls2c} are
marginally stable.

With the two mode approximation at $|v|=\infty$, we recapitulate the
parametric evolution of stationary states and eigenenergies along
$C(X)$. We assume that the system is initially in the ground state
$\Psi_1(x;g)$.

First, we revisit the case that the interparticle interaction is
weak enough, i.e., $g$ is smaller than the critical value $g_0$.
The stationary state
is localized at $|v|=\infty$, which is consistent with
Fig.~\ref{fig:nCyPsi_weak}.  The corresponding estimation of
eigenenergy at $|v|=\infty$ is given by Eq.~\eqref{eq:Elocalized}, which is
indicated in Fig.~\ref{fig:nCyE}.
Also, the stationary state at $|v|=\infty$ is stable, according to the
linear stability analysis. Hence the stationary state must be stable during
$C(X)$, which is also consistent with the numerical result that $C(X)$
induces the population inversion for smaller $g$
(Fig.~\ref{fig:pop2}).

Second, when the strength of the interparticle interaction $g$ exceeds the
critical value $g_0$, the localized and delocalized stationary
solutions coexist at the quench point: the degenerated eigenenergies of the
delocalized solutions $E_{\pm}(g)$ (Eq.~\eqref{eq:Edeloc})
is lower than the one of the localized
solution (Eq.~\eqref{eq:Elocalized}), as shown in Fig.~\ref{fig:nCyE}.
The delocalized nodeless solution $\Psi_+$
(Fig~\ref{fig:nCyPsi} ${\rm B}_{\pm}$) is connected with
the initial ground state $\Psi_1$ through $C_{\rm s}(0,\infty)$,
the former half of the whole cycle. After $\Psi_+$ is evolved
along the loop, the stationary solution arrives at the localized solution
$(\Psi_0, \Psi_1)=(1,0)$ at $|v|=\infty$
(see, Fig~\ref{fig:nCyPsi} ${\rm D}_{\pm}$).
After the completion of the
loop, the stationary solution becomes the another delocalized solution
$\Psi_-$ with a node (Fig~\ref{fig:nCyPsi}~${\rm F}$).
The latter half of the cycle $C_{\rm s}(-\infty,0)$
smoothly connects $\Psi_-$ with the first excited state $\Psi_2$. In this
sense, the parametric evolution of the stationary solution smoothly
deforms the ground state into the first excited state.
Meanwhile the adiabatic population inversion is hindered by the presence
of the loop structure due to the instability of the stationary state
at the lower branch of the loop, as explained in the previous section.

In Fig.~\ref{fig:pop2}, we indicate the
critical interparticle interaction strength $g_0$, which is estimated
by the two-mode approximation for each value of $X$ by vertical lines.
Hence we conclude that the two-mode approximation qualitatively describes
the breakdown of the population inversion.

\begin{figure}[htb]
  \centering
  \includegraphics[%
	width=7.0cm%
        ]{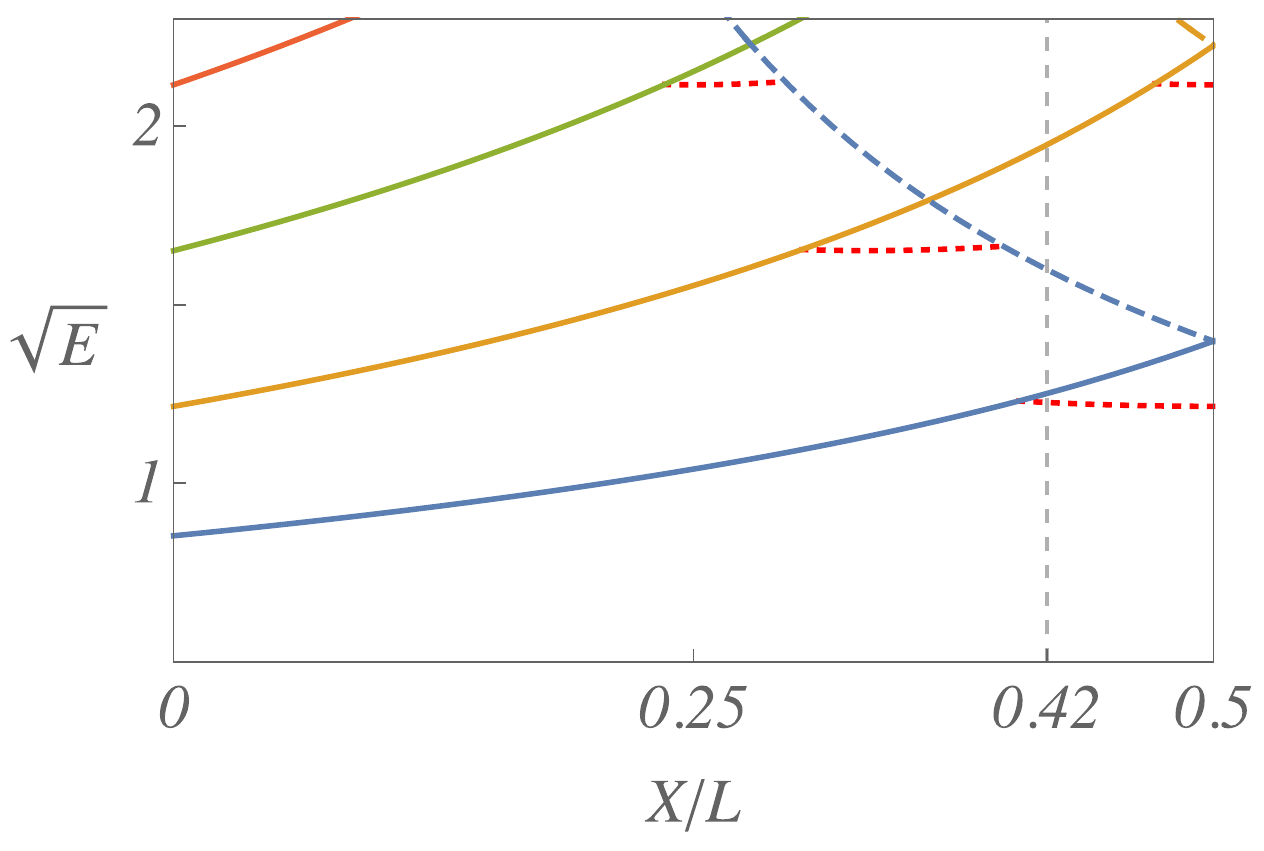}%
  \hspace*{1.00cm}%
  \caption{
    (Color online)
    $X$-dependence
    of stationary energies at the quench $|v|=\infty$ and $g=2$
    under the two-mode approximation.
    The solid (dashed) lines correspond to the stable stationary states localized at the right (left) side of the
    infinitely strong impurity.
    The dotted lines correspond to the delocalized stationary states that are marginally stable.
  }
  \label{fig:Cy_r_vs_E}
\end{figure}

We depict how the eigenenergies at the quench point depend on the
position $X$ of the sharp impurity in Fig.~\ref{fig:Cy_r_vs_E}, using
the two-mode approximation. This helps us to understand the cycle
$C(X)$ for a given value of $g$. For example, at $X=0.42 L$, the
ground branch connected with a loop whose section at $v=|\infty|$ has
two delocalized and a localized stationary states, which predicts the
breakdown of the population inversion to the first excited
states. Also, the first (the second) excited state at the initial
point of the cycle is connected with a localized stationary state,
which suggests that the population inversion to the second (the third)
excited state may be possible.

\section{Summary and outlook}
We have shown that the adiabatic
cyclic operation
with a
quench
$C(X)$ induces the population inversion of Bose
particles
described by the Gross-Pitaevskii equation
confined in a
quasi one-dimensional
box, if the Bose
particles are initially prepared to be in the ground state and the
strength $g$ of the interparticle interaction is weak enough. An
estimation of the critical value of $g$ where the population inversion
is broken is also shown. We find that these results are consistent
with our numerical investigation through the time-dependent
Gross-Pitaevskii equation.

We note that the time evolution generated by the almost-adiabatic
operation $C(X)$ can confine the system within a family of
stationary states in the weak interaction regime. Namely, the
adiabatic time evolution can be realized in spite of the presence of
the flip of the potential strength at the quench point.
This ``adiabatic'' cycle converts the ground state of Bose particles
into a nonequilibrium statonary state.
The present study offers an
example
of the subtle difference between
the adiabatic processes in thermodynamic systems and non-thermal,
mechanical systems.

We
believe
that the present result offers an experimentally feasible
method to produce a nonequilibrium stationary state. Application of
the acceleration of the adiabatic scheme to the present procedure
(e.g.,
Ref.~\cite{MartinezGaraot-PRL-111-213001,Masuda-PRL-113-063003})
should be also interesting.
We
note
that the preparation of condensates in a
quasi one-dimensional
box trap is experimentally achieved in
Ref.~\cite{Meyrath-PRA-71-041604}, which motives theoretical studies,
e.g., on solitonic excitations~\cite{Sciacca-PRA-95-013628}.
%
%
We also remark that the box trap may be useful to investigate quantum
information
processing through,
e.g. the Szilard engine in the quantum
regime~\cite{Kim-PRL-106-070401,Sodal-PRA-99-022121}. To extend these
works to many-body settings, our analysis on the
infinitely strong impurity
should
be applicable.

\section*{Acknowledgments}
A.T. wishes to thank Axel P\'erez-Obiol and Masaya Kunimi for
discussions.  This work has been supported by the Grant-in-Aid for
Scientific Research of MEXT, Japan (Grants No. 17K05584).

\appendix
\section{Linear stability analysis for the two-mode system\protect~\eqref{eq:nls2} at the quench point}
\label{sec:Bogoliubov}

We explain the linear stability analysis (Bogoliubov analysis) of the
stationary states at the quench point in $C(X)$.
To carry this out, the nonlinear Sch\"ordinger
equation~\eqref{eq:nls2} under the two-mode approximation is cast into
a nonlinear Bloch equation (Eq.~\eqref{eq:nb} below). The components
of Bloch vector $\bvec{S}=(S_x,S_y,S_z)$ are the expectation values of
Pauli matrices $\sigma_j$'s for a normalized state $(\Psi_0, \Psi_1)$,
e.g.,
$S_z = |\Psi_0|^2-|\Psi_1|^2$.

To find the time evolution equation of $\bvec{S}$, we first obtain a
matrix form of Eq.~\eqref{eq:nls2}. The system is described by an
effective nonlinear Hamiltonian
${H}_{2}=\Delta_{+}I+\Delta_{-}\sigma_z$,
where $I$ is the identity matrix, and
\begin{equation}
  \label{eq:defDelta}
  \Delta_{\pm}(\bvec{S})
  \equiv
  \frac{1}{2}\left\{(E_0+ gc_0\frac{1+S_z}{2})
    \pm(E_1+gc_1\frac{1-S_z}{2})\right\}
  .
\end{equation}
Hence $\bvec{S}$ experiences the effective magnetic field
\begin{equation}
  \label{eq:defB}
  \bvec{B}(\bvec{S})\equiv \Delta_{-}(\bvec{S})\bvec{e}_{z}
  ,
\end{equation}
where
$\bvec{e}_z\equiv(0,0,1)$.
Namely, $\bvec{S}$ obeys
the nonlinear Bloch equation
\begin{equation}
  \label{eq:nb}
  \frac{d}{dt}{\bvec{S}}=\bvec{S}\times\bvec{B}(\bvec{S})
  .
\end{equation}
A stationary state of Eq.~\eqref{eq:nls2} corresponds to a stationary
solution $\bvec{S}_*$ of Eq.~\eqref{eq:nb}, where
$\bvec{S}_*\times\bvec{B}(\bvec{S}_*)=0$ holds.

We proceed to the linear stability analysis for a stationary
solution~$\bvec{S}_*$ to examine a slightly perturbed Bloch vector
$\bvec{S}=\bvec{S}_* + \delta\bvec{S}$.
We expand $\delta\bvec{S}$,
using a orthogonal system $\bvec{e}_0\equiv \bvec{S}_*$,
$\bvec{e}_1\equiv \bvec{e}_y\left(=(0,1,0)\right)$, and
$\bvec{e}_2\equiv \bvec{e}_0\times\bvec{e}_1$,
as
\begin{equation}
  \label{eq:expandS}
  \delta\bvec{S}=\alpha_1\bvec{e}_1+\alpha_2\bvec{e}_2,
\end{equation}
where small coefficients $\alpha_j$'s are taken up to a first order
($j=1,2$).  The absence of the $\bvec{e}_0$ component in
Eq.~\eqref{eq:expandS} is consistent with the normalization condition
up to the first order.
The linearized equation for $\alpha_j$'s are
\begin{equation}
  \frac{d}{dt}\alpha_j
  =\bvec{e}_j\cdot
  \left\{
    \left[\delta{B}\right]\bvec{e}_z
    \times\bvec{S}_*
    + B_*\bvec{e}_z\times\delta\bvec{S}
  \right\}
  ,
\end{equation}
where
$B_*\equiv \Delta_{-}(\bvec{S}_*)$,
$\delta{B}=g(c_0+c_1)%
{S}_{*z}
\alpha_2/4$,
and ${S}_{*j}=\bvec{S}_{*}\cdot\bvec{e}_j$ ($j=x,y,z$),
from Eqs.~\eqref{eq:defB} and \eqref{eq:defDelta}.
Hence we obtain
\begin{equation}
  \frac{d}{dt}
  \begin{bmatrix}
    \alpha_1\\ \alpha_2
  \end{bmatrix}
  =
  M
  \begin{bmatrix}
    \alpha_1\\ \alpha_2
  \end{bmatrix}
\end{equation}
where
\begin{equation}
  \label{eq:defM}
  M \equiv
  \begin{bmatrix}
    0&
    -B_* {S}_{*z}+g\frac{c_0+c_1}{4}{S}_{*x}^2
    \\
    B_*{S}_{*z}&0
  \end{bmatrix}
  .
\end{equation}

We examine $M$ for each stationary solution $\bvec{S}_*$.
First, we examine the localized solutions $\bvec{S}_*=\pm\bvec{e}_z$,
where
\begin{equation}
  M =
  \pm B_*
  \begin{bmatrix}
    0& -1
    \\
    1&0
  \end{bmatrix}
  ,
\end{equation}
and $B_*$ is non-zero, except at the bifurcation point.
Since the eigenvalues of $M$ are purely imaginary,
the perturbation $\delta S$ evolves oscillatory,
and doesn't grow up exponentially in time.
Hence we conclude that the stationary solutions
$\bvec{S}_*=\pm\bvec{e}_z$ are linearly stable.

Second, we examine the delocalized solution Eq.~\eqref{eq:nls2c}, which
implies $B_*=0$. We find
\begin{equation}
  M =
  g\frac{c_0+c_1}{4}
  {S}_{*x}^2
  \begin{bmatrix}
    0&1\\
    0&0
  \end{bmatrix}
  ,
\end{equation}
which is non-zero, except at the bifurcation point.
Namely, $M$ has a non-trivial Jordan block and cannot be diagonalized.
In terms of dynamical systems, the stability of the
the delocalized solutions Eq.~\eqref{eq:nls2c} are marginally stable.
Although the perturbation $\delta S$ does not grows up exponentially fast,
it grows up linearly in $t$.



%


\end{document}